\documentstyle[12pt,amssymb,epsf]{article}

\renewcommand{\theequation}{\arabic{section}.\arabic{equation}}

\makeatletter
\def\slash#1{{\mathpalette\c@ncel{#1}}} 
\makeatother

\newcommand{\deriv}{\stackrel{\leftrightarrow}{D}}
\newcommand{\derleft}{\stackrel{\leftarrow}{D}}
\newcommand{\derright}{\stackrel{\rightarrow}{D}}
\newcommand{\dnabla}{\stackrel{\leftrightarrow}{\nabla}}
\newcommand\beq{\begin{eqnarray}}
\newcommand\eeq{\end{eqnarray}}
\newcommand\wt{\widetilde}
\newcommand\ud{\uparrow\downarrow}
\newcommand\du{\downarrow\uparrow}

\def\xslash{\rlap/{\mkern-1mu x}}

\renewcommand{\thefootnote}{\fnsymbol{footnote}}

\begin{document}


\begin{titlepage}
\begin{flushright}
\begin{tabular}{l}
CERN--TH/98--333\\
NORDITA--98--62--HE\\
hep-ph/9810475
\end{tabular}
\end{flushright}
\vskip0.5cm
\begin{center}
  {\Large \bf 
             Higher Twist Distribution Amplitudes of Vector Mesons in
             QCD: Twist-4  Distributions and\\ Meson Mass Corrections
  \\}

\vspace{1cm}
{\sc Patricia~Ball}${}^{1,}$\footnote{E-mail: Patricia.Ball@cern.ch} and
{\sc V.M.~Braun}${}^{2,}$\footnote{E-mail: vbraun@nordita.dk}
\\[0.5cm]
\vspace*{0.1cm} ${}^1${\it CERN--TH, CH--1211 Geneva 23, Switzerland
}\\[0.3cm]
\vspace*{0.1cm} ${}^2${\it NORDITA, Blegdamsvej 17, DK-2100 Copenhagen,
Denmark}
\\[1.3cm]

  \vfill

  {\large\bf Abstract\\[10pt]} \parbox[t]{\textwidth}{ 
  We present a systematic study of  twist-4 light-cone distribution
  amplitudes of vector mesons in QCD, which is based on conformal expansion.
  The structure of meson mass corrections is studied in detail.
  A complete set of distribution amplitudes is constructed, which satisfies 
  all (exact) equations of motion and constraints from conformal expansion. 
  Nonperturbative input parameters are estimated from QCD sum rules.
  Our study suggests that meson mass corrections may present a dominant 
  source of higher twist effects in exclusive processes.
                                   }
  \vskip1cm 
{\em Submitted to Nuclear Physics B}\\[1cm]
\end{center}
\end{titlepage}

\setcounter{footnote}{0}
\renewcommand{\thefootnote}{\arabic{footnote}}

\section{Introduction}
\setcounter{equation}{0}

The notion of distribution amplitudes refers to 
momentum fraction distributions of partons in a meson, in a particular 
Fock state, with a fixed number of constituents. For the minimal number of 
constituents, the distribution amplitude $\phi$ is related to the 
meson's Bethe-Salpeter wave function $\phi_{BS}$ by
\begin{equation}
  \phi(x) \sim \int^{|k_\perp| < \mu} \!\!d^2 k_\perp\,
\phi_{BS}(x,k_\perp).
\end{equation}   
The standard approach to distribution amplitudes, which is due to Brodsky
and Lepage~\cite{BLreport}, considers the hadron's parton decomposition
in the infinite momentum frame. A conceptually different, but
mathematically 
equivalent formalism is the light-cone quantization~\cite{LCQ}.
Either way, power suppressed contributions to exclusive processes in QCD,
which are commonly referred to as higher twist corrections, are thought to 
originate from three different sources:
\begin{itemize}
\item contributions of ``bad'' components in the wave function and 
      in particular of components with ``wrong'' spin projection;
\item contributions of transverse motion of quarks (antiquarks) in the 
      leading twist components, given for instance by integrals as above with 
      additional factors of $k_\perp^2$;
\item contributions of higher Fock states with additional gluons and/or
      quark-antiquark pairs. 
\end{itemize} 

In this paper we continue the systematic study of higher twist 
light-cone distribution amplitudes started in Ref.~\cite{BBKT}.
In particular, we extend the analysis of \cite{BBKT} to include
twist-4 distribution amplitudes and, most significantly, 
 meson mass corrections. A preliminary account 
of some of our results has been reported in  \cite{handbook}.  

Following \cite{BBKT}, we define 
light-cone distribution amplitudes as meson-to-vacuum transition matrix 
elements of nonlocal gauge invariant light-cone operators. This formalism 
is perhaps less intuitive than the infinite momentum frame formulation,
but it is more convenient for the study of higher twist distributions 
as it is Lorentz and gauge invariant.  
It  allows  all equations of motion 
to be solved 
explicitly, relating different higher twist distributions to one another.   
We will find that, much like in the twist-3 case \cite{BBKT},
 all dynamical degrees of freedom are those describing 
contributions of higher Fock states, while all other higher twist effects 
are given in terms of the latter without any free parameters.

A systematic study of meson mass corrections presents
the principal new contribution of this work. 
By counting dimensions, for any exclusive observable involving 
a large momentum transfer $Q$, power suppressed  
higher twist corrections have the generic structure
$$ 
\frac{1}{Q^2} \Big[m^2\cdot \langle\!\langle O^{(2)}\rangle\!\rangle
+ m \cdot\langle\!\langle O^{(3)}\rangle\!\rangle
   + \langle\!\langle O^{(4)}\rangle\!\rangle\Big].
$$
Here $m$ is the meson mass,
$\langle\!\langle O^{(2)}\rangle\!\rangle$ and 
$\langle\!\langle O^{(3)}\rangle\!\rangle$ and 
$\langle\!\langle O^{(4)}\rangle\!\rangle$ are reduced matrix elements
of twist-2, twist-3 and twist-4 operators, which have 
dimension 0, 1 and 2, respectively.
The terms $\sim m^2$ do not involve any new dynamical information 
about the meson structure as compared to the leading twist terms, and 
are usually  referred to as ``kinematic'' power corrections.
The structure of such kinematic corrections is well known 
for deep-inelastic lepton-hadron scattering, in which case  they can
be absorbed into a redefinition of the scaling variable~\cite{N73}.
The crucial observation leading to this ``Nachtmann scaling''
is that hadron mass corrections (``target mass corrections'' in this 
context) arise exclusively from the definition of the relevant 
leading twist matrix elements and do not involve new (higher twist) operators.
This simplification does not hold in exclusive processes because there
are additional
contributions of operators containing total derivatives.  
Specifically, to twist-4 accuracy, in addition to Nachtmann's
corrections, there are also contributions of operators of type
$$\partial^2 O^{(2)}_{\mu_1\mu_2\ldots\mu_n}$$
and
$$\partial_{\mu_1}O^{(2)}_{\mu_1\mu_2\ldots\mu_n},$$ 
where $O^{(2)}$ is a leading twist operator.
We find that contributions of the first type can be taken into 
account  consistently for all moments, while contributions of the
second type are more complicated and can be unravelled only order by 
order in the conformal expansion. 

The outline of this paper is as follows:
 definitions of and notations for distribution amplitudes are 
presented in Sec.~2 together with general remarks about  
specific features  of the operator product expansion (OPE) for 
exclusive processes and about conformal expansion.
Section 3 gives a general discussion of meson mass corrections 
for a simple example.
The subsequent Secs.~4 and 5
contain a detailed derivation of chiral-even and chiral-odd 
distribution amplitudes, respectively. We take into  account 
contributions of the leading and next-to-leading conformal spin 
and derive a self-consistent approximation for the 
distribution amplitudes, which respects the exact QCD equations of 
motion. The chiral-even and chiral-odd
asymptotic distribution amplitudes  
 involve three nonperturbative parameters, and 
four additional  parameters are required  for the description of
the leading corrections. The corresponding estimates are worked out 
using the QCD sum rule approach~\cite{SVZ}. On the basis of these estimates,
we suggest  that higher twist effects in exclusive processes are 
in many cases  dominated by meson mass corrections alone.   
The final Sec.~6 contains a summary and conclusions. The paper also
 contains three appendices in which we derive equations of motion for
 nonlocal operators, and derive and estimate the independent
 nonperturbative parameters that enter the twist-4 distributions
 discussed here.

Throughout this paper we denote the meson momentum by $P_\mu$  
and introduce light-like vectors $p$ and $z$ such that 
\begin{equation}
p_\mu = P_\mu-\frac{1}{2}z_\mu \frac{m^2_\rho}{pz}.
\label{smallp}
\end{equation}  
The meson polarization vector $e^{(\lambda)}_\mu$ is decomposed into 
projections onto the two light-like vectors and the orthogonal plane 
as
\begin{equation}
 e^{(\lambda)}_\mu = \frac{(e^{(\lambda)} z)}{pz}
\left( p_\mu -\frac{m^2_\rho}{2pz} z_\mu \right)+e^{(\lambda)}_{\perp\mu}. 
\label{polv}
\end{equation} 
Some useful scalar products are
\begin{eqnarray}
  z\cdot P = z\cdot p &=& \sqrt{(x\cdot P)^2-x^2m_\rho^2},
\nonumber\\
  p\cdot e^{(\lambda)} &=& -\frac{m^2_\rho}{2 pz}\, z\cdot e^{(\lambda)},
\nonumber\\
 e^{(\lambda)}\cdot z &=& e^{(\lambda)}\cdot x.
\end{eqnarray}
We also need the projector onto the directions orthogonal to $p$ and $z$:
\begin{equation}
       g^\perp_{\mu\nu} = g_{\mu\nu} -\frac{1}{pz}(p_\mu z_\nu+ p_\nu z_\mu), 
\end{equation}
and will use the notations
\begin{equation}
    a.\equiv a_\mu z^\mu, \qquad a_\ast \equiv a_\mu p^\mu/(pz),
\end{equation}
for an arbitrary Lorentz vector $a_\mu$.

We use the standard Bjorken--Drell 
convention \cite{BD65} for the metric and the Dirac matrices; in particular
$\gamma_{5} = i \gamma^{0} \gamma^{1} \gamma^{2} \gamma^{3}$,
and the Levi--Civita tensor $\epsilon_{\mu \nu \lambda \sigma}$
is defined as the totally antisymmetric tensor with $\epsilon_{0123} = 1$.
The covariant derivative is defined as 
$D_{\mu} \equiv \overrightarrow{D}_\mu= \partial_{\mu} - igA_{\mu}$, 
and we also use the notation
$\overleftarrow{D}_{\mu} = \overleftarrow\partial_{\mu} 
+ig A_{\mu}$ in later sections. The dual gluon field strength
tensor is defined as $\widetilde{G}_{\mu\nu} =
\frac{1}{2}\epsilon_{\mu\nu \rho\sigma} G^{\rho\sigma}$.

\section{General Framework}
\setcounter{equation}{0}

Amplitudes of light-cone-dominated processes involving vector mesons
can be expressed in terms of matrix elements of gauge invariant
nonlocal operators sandwiched between the vacuum and the vector meson
state, e.g.\ a matrix element over a two-particle operator,
\begin{equation}
    \langle 0|\bar u(x) \Gamma [x,-x] d(-x)|\rho^-(P,\lambda)\rangle,
\label{eq:1}
\end{equation}
where $\Gamma$ is a generic Dirac matrix structure
and we use the notation 
$[x,y]$ for the path-ordered gauge factor along the straight line 
connecting the points $x$ and $y$:
\begin{equation}
[x,y] ={\rm P}\exp\left[ig\!\!\int_0^1\!\! dt\,(x-y)_\mu
  A^\mu(tx+(1-t)y)\right].
\label{Pexp}
\end{equation}
To simplify the notation,  we will explicitly consider  
charged $\rho$ mesons; the distribution amplitudes  of $\rho^0$ 
can  be obtained by choosing appropriate isospin currents.

The asymptotic expansion of exclusive amplitudes in powers of 
large momentum transfer corresponds to the expansion
of amplitudes like (\ref{eq:1}) in powers of the deviation 
from the light-cone $x^2= 0$. As always in a quantum field theory,
such an 
expansion generates divergences and has to be understood as an 
OPE in terms of renormalized light-cone nonlocal
operators whose matrix elements define meson distribution amplitudes 
of increasing twist. To  leading logarithmic accuracy,
the coefficient functions are just taken at tree-level, and the 
distributions have to be evaluated at the scale $\mu^2 \sim x^{-2}$.
In this section we present the necessary expansions 
and introduce a complete set of meson distribution amplitudes
to twist-4 accuracy. This set is, in fact, overcomplete, and different
distributions are related to one another via the QCD equations of motion, as
 detailed in later sections. 
 
\subsection{Chiral-Even Distribution Amplitudes}

We start with the matrix elements involving an odd number 
of $\gamma$ matrices, which we refer to as chiral-even in what follows.  
For the vector and axial vector operators 
the light-cone expansion to twist-4 accuracy reads:
\begin{eqnarray}
\langle 0|\bar u(x) \gamma_\mu d(-x)|\rho^-(P,\lambda)\rangle 
 &=& f_\rho m_\rho \Bigg\{
\frac{e^{(\lambda)}x}{Px}\, P_\mu \int_0^1 du \,e^{i\xi Px}
\Big[\phi_\parallel(u)
+\frac{m^2_\rho x^2}{4}  {\Bbb A}(u)\Big]
\nonumber\\
&&{}+\left(e^{(\lambda)}_\mu-P_\mu\frac{e^{(\lambda)}x}{Px}\right)
\int_0^1 du\, e^{i\xi Px} \,{\Bbb B}(u)
\nonumber\\&&{}
-\frac{1}{2}x_\mu \frac{e^{(\lambda)}x}{(Px)^2} m^2_\rho \int_0^1 du 
\, e^{i\xi Px} {\Bbb C} (u)
\Bigg\},
\label{eq:OPEvector}\\
\langle 0|\bar u(x) \gamma_{\mu} \gamma_{5} 
d(-x)|\rho^-(P,\lambda)\rangle 
&=& \frac{1}{2}\left(f_{\rho} - f_{\rho}^{T}
\frac{m_{u} + m_{d}}{m_{\rho}}\right)
m_{\rho} \epsilon_{\mu}^{\phantom{\mu}\nu \alpha \beta}
e^{(\lambda)}_{\nu} P_{\alpha} x_{\beta}
\int_{0}^{1} \!du\, e^{i \xi P  x} {\Bbb D}(u).\nonumber\\[-20pt]
\label{eq:OPEaxial}
\end{eqnarray}
Notice that in order to calculate exclusive amplitudes to
 $O(1/Q^2)$ accuracy, terms of $O(x^2)$ have to be kept 
in the vector matrix element, but can be neglected in the axial vector
 one. 
For brevity, here and below we do not show gauge factors between the 
quark and the antiquark fields; we also use the short-hand notation
$$\xi = 2u-1.$$ The vector and tensor  decay constants 
$f_\rho$ and $f_\rho^T$ are defined, as usual, as
\begin{eqnarray}
\langle 0|\bar u(0) \gamma_{\mu}
d(0)|\rho^-(P,\lambda)\rangle & = & f_{\rho}m_{\rho}
e^{(\lambda)}_{\mu},
\label{eq:fr}\\
\langle 0|\bar u(0) \sigma_{\mu \nu} 
d(0)|\rho^-(P,\lambda)\rangle &=& i f_{\rho}^{T}
(e_{\mu}^{(\lambda)}P_{\nu} - e_{\nu}^{(\lambda)}P_{\mu}).
\label{eq:frp}
\end{eqnarray}
The coupling $f_\rho^T$ is scale-dependent, with
\begin{equation}
f_\rho^T(Q^2) =
f_\rho^T(\mu^2)\,\left(\frac{\alpha_s(Q^2)}{\alpha_s(\mu^2)}
\right)^{C_F/b},\label{eq:fTscaling}
\end{equation}
with the standard notation $C_F = (N_c^2-1)/(2N_c)$ and $b = (11N_c-2 n_f)/3$.

The expansions in (\ref{eq:OPEvector}), (\ref{eq:OPEaxial})
involve several Lorentz invariant amplitudes, which we have 
to interpret in terms of meson parton distributions.
Definitions of the latter involve nonlocal operators at 
strictly light-like separations and can most conveniently be 
written using the light-cone variables (\ref{smallp}), and  
for longitudinal and transverse meson polarizations separately.

Following \cite{BBKT}, we define chiral-even 
two-particle distribution amplitudes of the $\rho$ meson as
\begin{eqnarray}
\lefteqn{\langle 0|\bar u(z) \gamma_{\mu} d(-z)|\rho^-(P,\lambda)\rangle 
 = f_{\rho} m_{\rho} \left[ p_{\mu}
\frac{e^{(\lambda)} z}{p  z}
\int_{0}^{1} \!du\, e^{i \xi p  z} \phi_{\parallel}(u, \mu^{2}) \right. 
}\hspace*{0.6cm}\nonumber \\
&+&\left. e^{(\lambda)}_{\perp \mu}
\int_{0}^{1} \!du\, e^{i \xi p  z} g_{\perp}^{(v)}(u, \mu^{2}) 
- \frac{1}{2}z_{\mu}
\frac{e^{(\lambda)} z }{(p  z)^{2}} m_{\rho}^{2}
\int_{0}^{1} \!du\, e^{i \xi p  z} g_{3}(u, \mu^{2})
\right],\hspace*{0.4cm}
\label{eq:vda}\\
\lefteqn{\langle 0|\bar u(z) \gamma_{\mu} \gamma_{5} 
d(-z)|\rho^-(P,\lambda)\rangle=}\hspace*{0.6cm}\nonumber\\ 
&=& \frac{1}{2}\left(f_{\rho} - f_{\rho}^{T}
\frac{m_{u} + m_{d}}{m_{\rho}}\right)
m_{\rho} \epsilon_{\mu}^{\phantom{\mu}\nu \alpha \beta}
e^{(\lambda)}_{\perp \nu} p_{\alpha} z_{\beta}
\int_{0}^{1} \!du\, e^{i \xi p  z} g^{(a)}_{\perp}(u, \mu^{2}).
\label{eq:avda}
\end{eqnarray} 
The distribution amplitude $\phi_\parallel$ is of twist-2,
$g_\perp^{(v)}$ and $g_\perp^{(a)}$ of twist-3 and $g_3$ of twist-4. 
All four functions $\phi=\{\phi_\parallel,
g_\perp^{(v)},g_\perp^{(a)},g_3\}$ are normalized as
\begin{equation}
\int_0^1\!du\, \phi(u) =1,
\label{eq:norm}
\end{equation}
which can be checked by comparing the two sides
of the  defining equations in the limit $z_\mu\to 0$ and using the
equations of motion.

Comparing (\ref{eq:vda}), (\ref{eq:avda}) with the light-cone expansions
in (\ref{eq:OPEvector}), (\ref{eq:OPEaxial}), we easily find
\begin{eqnarray}
{\Bbb B}(u) &=& g^{(v)}_\perp(u),    
\nonumber\\
{\Bbb C}(u) &=& g_3(u)+\phi_\parallel(u) -2 g^{(v)}_\perp(u),    
\nonumber\\
{\Bbb D}(u) &=& g^{(a)}_\perp(u),    
\end{eqnarray}
which is nothing but the tree-level OPE of the invariant amplitudes
${\Bbb B},{\Bbb C},{\Bbb D}$ in terms of meson distribution amplitudes.
We will find (see also \cite{BBKT}) that all higher twist two-particle 
distribution amplitudes do not present genuine independent degrees
of freedom, but can be expressed in terms of three-particle 
distribution amplitudes. The same analysis will allow us to calculate the 
remaining invariant amplitude ${\Bbb A}$, which accounts for the 
transverse momentum distribution in the valence component of the 
wave function.   

\begin{table}
\addtolength{\arraycolsep}{3pt}
\renewcommand{\arraystretch}{1.4}
$$
\begin{array}{|c|lcc|lc|lcc|}
\hline
{\rm Twist} &(\mu\nu\alpha) 
& \bar\psi \widetilde{G}_{\mu\nu}\gamma_\alpha\gamma_5
\psi & \bar\psi G_{\mu\nu}\gamma_\alpha \psi 
&(\mu\nu\alpha\beta) & \bar\psi G_{\mu\nu} \sigma_{\alpha\beta}\psi 
&(\mu\nu) & \bar\psi G_{\mu\nu} \psi & \bar\psi \widetilde{G}_{\mu\nu}\gamma_5
\psi\\ \hline
3 & \cdot\perp \cdot & {\cal A} & {\cal V} & \cdot\perp
   \cdot\perp & {\cal T}&&&\\ \hline
4 & \cdot\perp\perp & \widetilde{\Phi} & \Phi & 
\perp\perp\cdot\!\perp & T_1^{(4)}
& \cdot\perp & S & \widetilde{S}\\
 & \cdot * \cdot & \widetilde{\Psi} & \Psi & \cdot\perp
 \perp\perp & T_2^{(4)}&&&\\ 
& & & & \cdot * \cdot\perp & T_3^{(4)}&&&\\
& & & & \cdot\perp\! \cdot\, * & T_4^{(4)}&&&\\\hline
\end{array}
$$
\addtolength{\arraycolsep}{-3pt}
\renewcommand{\arraystretch}{1}
\caption[]{
Identification of three-particle distribution amplitudes with 
projections onto different light-cone components of the 
nonlocal operators. For example, $\cdot\perp\perp$ refers to
$\bar\psi \widetilde{G}_{\cdot\perp}\gamma_\perp\gamma_5\psi$.
}
\end{table}

Three-particle chiral-even distributions are rather numerous and can be 
defined by the following matrix elements:
\begin{eqnarray}
\langle 0|\bar u(z) g\widetilde G_{\mu\nu}(vz)\gamma_\alpha\gamma_5 
  d(-z)|\rho^-(P,\lambda)\rangle & = &
  f_\rho m_\rho p_\alpha[p_\nu e^{(\lambda)}_{\perp\mu}
 -p_\mu e^{(\lambda)}_{\perp\nu}]{\cal A}(v,pz)
\nonumber\\ &
 & {}+f_\rho m_\rho^3\frac{e^{(\lambda)} z}{pz}
[p_\mu g^\perp_{\alpha\nu}-p_\nu g^\perp_{\alpha\mu}] \widetilde\Phi(v,pz)
\nonumber\\& & {}+
 f_\rho m_\rho^3\frac{e^{(\lambda)} z}{(pz)^2}
p_\alpha [p_\mu z_\nu - p_\nu z_\mu] \widetilde\Psi(v,pz),\label{eq:even1}\\
\langle 0|\bar u(z) g G_{\mu\nu}(vz)i\gamma_\alpha 
  d(-z)|\rho^-(P)\rangle &=&
  f_\rho m_\rho p_\alpha[p_\nu e^{(\lambda)}_{\perp\mu} 
  - p_\mu e^{(\lambda)}_{\perp\nu}]{\cal V}(v,pz)
\nonumber\\&&
{}+ f_\rho m_\rho^3\frac{e^{(\lambda)} z}{pz}
[p_\mu g^\perp_{\alpha\nu} - p_\nu g^\perp_{\alpha\mu}] \Phi(v,pz)
\nonumber\\& & {}+f_\rho m_\rho^3\frac{e^{(\lambda)} z}{(pz)^2}
p_\alpha [p_\mu z_\nu - p_\nu z_\mu] \Psi(v,pz),\label{eq:even2}
\end{eqnarray}
where 
\begin{equation}
   {\cal A}(v,pz) =\int {\cal D}\underline{\alpha} 
e^{-ipz(\alpha_u-\alpha_d+v\alpha_g)}{\cal A}(\underline{\alpha}),
\end{equation}
etc., and $\underline{\alpha}$ is the set of three momentum fractions
$\underline{\alpha}=\{\alpha_d,\alpha_u,\alpha_g\}$.
 The integration measure is defined as 
\begin{equation}
 \int {\cal D}\underline{\alpha} \equiv \int_0^1 d\alpha_d
  \int_0^1 d\alpha_u\int_0^1 d\alpha_g \,\delta\left(1-\sum \alpha_i\right).
\label{eq:measure}
\end{equation}
The distribution amplitudes ${\cal V}$ and ${\cal A}$ are of twist-3,
while the rest is of twist-4; we have not shown further Lorentz structures 
corresponding to twist-5 contributions\footnote{Note that we use  
a different normalization of three-particle twist-3 distributions
compared to \cite{BBKT}.}.
Different distribution amplitudes can be separated by projecting onto
 particular light-cone components, as summarized in Table~1.

For completeness, let us mention that also four-particle twist-4
distribution amplitudes exist, corresponding to contributions of 
Fock states with two gluons or an additional $q \bar q$ pair, of 
type
$$
  \bar\psi \gamma_. (\gamma_5)\psi\,
  \bar\psi \gamma_. (\gamma_5)\psi,
~~~~  \bar\psi \,G_{.\perp}\,G_{.\perp} \gamma_. \psi.
$$  
Such distributions will not be considered in this paper for two
reasons: first, it is well known \cite{SV82} that four-particle 
twist-4 operators do not allow the    factorization of  vacuum condensates
such as $\langle \bar\psi \psi\rangle$, $\langle G^2\rangle$. 
Because of this,  their matrix elements cannot be estimated reliably by
existing methods (e.g.\ QCD sum rules), although they are
generally  expected to be small.
Second, and more importantly, the four-particle 
distributions decouple from the QCD equations of motion in the two lowest 
conformal partial waves. To this accuracy, therefore, it is consistent
to put them to zero. Vice versa, nonvanishing four-particle 
distributions necessitate the inclusion of higher conformal spin 
corrections to distributions with less particles, which are beyond 
the approximation adopted in this paper.    

\subsection{Chiral-Odd Distribution Amplitudes}

For chiral-odd operators involving $\sigma_{\mu\nu}$ and {\bf 1}, the
light-cone expansion to twist-4 accuracy reads:
\begin{eqnarray}
\langle 0|\bar u(x) \sigma_{\mu \nu}
d(-x)|\rho^-(P,\lambda)\rangle & = &
 i f_{\rho}^{T} \left[ (e^{(\lambda)}_{\mu}P_\nu -
e^{(\lambda)}_{\nu}P_\mu )
\int_{0}^{1} \!du\, e^{i \xi P x}
\Bigg[\phi_{\perp}(u) +\frac{m_\rho^2x^2}{4} {\Bbb A}_T(u)\Bigg] \right.
\nonumber \\
& &{}+ (P_\mu x_\nu - P_\nu x_\mu )
\frac{e^{(\lambda)} x}{(P x)^{2}}
m_{\rho}^{2}
\int_{0}^{1} \!du\, e^{i \xi P x} {\Bbb B}_T (u)
\nonumber \\
& & \left.{}+ \frac{1}{2}
(e^{(\lambda)}_{ \mu} x_\nu -e^{(\lambda)}_{ \nu} x_\mu)
\frac{m_{\rho}^{2}}{P  x}
\int_{0}^{1} \!du\, e^{i \xi P x} {\Bbb C}_T(u) \right],
\label{eq:OPE2}\\
\langle 0 | \bar u(x) d(-x) | \rho^-(P,\lambda)\rangle & = & -i
\left(f_\rho^T - f_\rho\,\frac{m_u+m_d}{m_\rho}\right)
\left(e^{(\lambda)} x\right) m_\rho^2 \int_0^1 du\, e^{i\xi Px} {\Bbb D}_T(u).
\label{eq:OPEx}
\end{eqnarray}
The couplings $f_\rho$ and $f_\rho^T$ are defined in (\ref{eq:fr}) and
(\ref{eq:frp}). 

The corresponding distribution amplitudes on the light-cone are
defined as
\begin{eqnarray}
\langle 0|\bar u(z) \sigma_{\mu \nu}
d(-z)|\rho^-(P,\lambda)\rangle
& = & i f_{\rho}^{T} \left[ ( e^{(\lambda)}_{\perp \mu}p_\nu -
e^{(\lambda)}_{\perp \nu}p_\mu )
\int_{0}^{1} \!du\, e^{i \xi p  z} \phi_{\perp}(u, \mu^{2}) \right.
\nonumber \\
& &{}+ (p_\mu z_\nu - p_\nu z_\mu )
\frac{e^{(\lambda)}  z}{(p z)^{2}}
m_{\rho}^{2}
\int_{0}^{1} \!du\, e^{i \xi p  z} h_\parallel^{(t)} (u, \mu^{2})
\nonumber \\
& & \left.{}+ \frac{1}{2}
(e^{(\lambda)}_{\perp \mu} z_\nu -e^{(\lambda)}_{\perp \nu} z_\mu)
\frac{m_{\rho}^{2}}{p  z}
\int_{0}^{1} \!du\, e^{i \xi p  z} h_{3}(u, \mu^{2}) \right],
\label{eq:tda}\\
\langle 0|\bar u(z)
d(-z)|\rho^-(P,\lambda)\rangle & = & 
 {} -i \left(f_{\rho}^{T} - f_{\rho}\frac{m_{u} + m_{d}}{m_{\rho}}
\right)(e^{(\lambda)} z) m_{\rho}^{2}
\int_{0}^{1} \!du\, e^{i \xi p  z} h_\parallel^{(s)}(u, \mu^{2}).
\nonumber\\[-20pt]\label{eq:sda}
\end{eqnarray}
The distribution amplitude $\phi_\perp$ is of twist-2,
$h_\parallel^{(s,t)}$ are of twist-3 and $h_3$ is of twist-4. All four
functions $\phi = \{\phi_\perp,h_\parallel^{(s,t)},h_3\}$ are
normalized as 
$$
\int_0^1 du\, \phi(u) = 1.
$$
Comparing (\ref{eq:tda}) and (\ref{eq:sda}) with the light-cone expansion
(\ref{eq:OPE2}) and (\ref{eq:OPEx}), we easily find
\begin{eqnarray}
   {\Bbb B}_T(u) &=& h_\parallel^{(t)}(u) -\frac{1}{2}\phi_\perp(u)-
              \frac{1}{2} h_3(u),
\nonumber\\
   {\Bbb C}_T(u) &=& h_3(u)-\phi_\perp(u).
\end{eqnarray}
As for chiral-even distribution amplitudes, only the twist-2
distribution $\phi_\perp$ represents genuinely independent degrees of
freedom, the others can be expressed in terms of three-particle
distribution amplitudes. 

The three-particle distribution amplitudes are even more numerous than 
in the chiral-even case and can be defined as:
\begin{eqnarray}
\lefteqn{\langle 0|\bar u(z) \sigma_{\alpha\beta}
         gG_{\mu\nu}(vz)
         d(-z)|\rho^-(P,\lambda)\rangle \ =}\hspace*{1.6cm}\nonumber\\
&=& f_{\rho}^T m_{\rho}^2 \frac{e^{(\lambda)} z }{2 (p  z)}
    [ p_\alpha p_\mu g^\perp_{\beta\nu}
     -p_\beta p_\mu g^\perp_{\alpha\nu}
     -p_\alpha p_\nu g^\perp_{\beta\mu}
     +p_\beta p_\nu g^\perp_{\alpha\mu} ]
      {\cal T}(v,pz)
\nonumber\\
&&{}+ f_{\rho}^T m_{\rho}^2
    [ p_\alpha e^{(\lambda)}_{\perp\mu}g^\perp_{\beta\nu}
     -p_\beta e^{(\lambda)}_{\perp\mu}g^\perp_{\alpha\nu}
     -p_\alpha e^{(\lambda)}_{\perp\nu}g^\perp_{\beta\mu}
     +p_\beta e^{(\lambda)}_{\perp\nu}g^\perp_{\alpha\mu} ]
      T_1^{(4)}(v,pz)
\nonumber\\
&&{}+ f_{\rho}^T m_{\rho}^2
    [ p_\mu e^{(\lambda)}_{\perp\alpha}g^\perp_{\beta\nu}
     -p_\mu e^{(\lambda)}_{\perp\beta}g^\perp_{\alpha\nu}
     -p_\nu e^{(\lambda)}_{\perp\alpha}g^\perp_{\beta\mu}
     +p_\nu e^{(\lambda)}_{\perp\beta}g^\perp_{\alpha\mu} ]
      T_2^{(4)}(v,pz)
\nonumber\\
&&{}+ \frac{f_{\rho}^T m_{\rho}^2}{pz}
    [ p_\alpha p_\mu e^{(\lambda)}_{\perp\beta}z_\nu
     -p_\beta p_\mu e^{(\lambda)}_{\perp\alpha}z_\nu
     -p_\alpha p_\nu e^{(\lambda)}_{\perp\beta}z_\mu
     +p_\beta p_\nu e^{(\lambda)}_{\perp\alpha}z_\mu ]
      T_3^{(4)}(v,pz)
\nonumber\\
&&{}+ \frac{f_{\rho}^T m_{\rho}^2}{pz}
    [ p_\alpha p_\mu e^{(\lambda)}_{\perp\nu}z_\beta
     -p_\beta p_\mu e^{(\lambda)}_{\perp\nu}z_\alpha
     -p_\alpha p_\nu e^{(\lambda)}_{\perp\mu}z_\beta
     +p_\beta p_\nu e^{(\lambda)}_{\perp\mu}z_\alpha ]
      T_4^{(4)}(v,pz),\hspace*{1cm}
\label{eq:T3}\\
\lefteqn{\langle 0|\bar u(z)
         gG_{\mu\nu}(vz)
         d(-z)|\rho^-(P,\lambda)\rangle
\ =\ i f_{\rho}^T m_{\rho}^2
 [e^{(\lambda)}_{\perp\mu}p_\nu-e^{(\lambda)}_{\perp\nu}p_\mu] S(v,pz),}
\hspace*{1.6cm}\nonumber\\
\lefteqn{\langle 0|\bar u(z)
         ig\widetilde G_{\mu\nu}(vz)\gamma_5
         d(-z)|\rho^-(P,\lambda)\rangle
\ =\ i f_{\rho}^T m_{\rho}^2
 [e^{(\lambda)}_{\perp\mu}p_\nu-e^{(\lambda)}_{\perp\nu}p_\mu]
  \widetilde S(v,pz).}\hspace*{1.6cm}\label{eq:2.21}
\end{eqnarray}
Of these seven amplitudes, ${\cal T}$ is of twist-3 and the other six
of twist-4; higher twist terms are suppressed. The relation of these
distribution amplitudes to specific light-cone projections of the
matrix elements was made explicit in Table~1.

Also in this case there exist four-particle twist-4 distribution
amplitudes which we do not consider for the reasons mentioned at the
end of Sec.~2.1.

\subsection{Conformal Partial Wave Expansion}

Conformal partial wave expansion in 
QCD~\cite{B+,Makeenko,O82,BF90,Mueller}
parallels the partial wave expansion
of wave functions in standard quantum mechanics,
which allows the separation of the dependence on angular 
coordinates from that on radial ones. The basic idea is to write 
distribution amplitudes as a sum of contributions from different 
conformal spins. For a given spin, the dependence on the momentum 
fractions is fixed by the symmetry. To specify the function, 
one has to fix the coefficients in this expansion at some scale;
 conformal invariance of the QCD Lagrangian then guarantees that there 
is no mixing between contributions of different spin to leading logarithmic 
accuracy. For leading twist distributions the mixing matrix becomes 
diagonal in the conformal basis and
the anomalous dimensions are ordered with spin.  
Thus, only the first few ``harmonics'' contribute at sufficiently large scales 
(for sufficiently hard processes).
  
For higher twist distributions, the use of the conformal basis offers 
the crucial  advantage of ``diagonalizing'' the equations of motion:
since conformal transformations commute
with the QCD equations of motion, the corresponding constraints
 can be solved order by order in the conformal expansion.
Note that relations between different distributions obtained in this way
are exact: despite the fact that conformal symmetry is broken by quantum 
corrections, equations of motion are not renormalized and remain the same as 
in free theory.  

The general procedure to construct the conformal expansion
for arbitrary multi-particle distributions was developed in \cite{O82,BF90}.
To this end, each constituent field has to be decomposed (using projection 
operators, if necessary) into components  with fixed (Lorentz) spin 
projection onto the light-cone.
 
Each such component corresponds to a so-called quasi-primary field in 
the language of 
conformal field theories, and has conformal spin 
\begin{equation}
  j=\frac{1}{2}\,(l+s),
\label{eq:cspin}
\end{equation}
where $l$ is the canonical dimension  and $s$ the (Lorentz) spin 
projection. In particular, $l=3/2$ for quarks and $l=2$ for gluons. 
The  quark field is decomposed as $\psi_+ \equiv 
(1/2)\slash{z}\slash{p}\psi$ and $\psi_-=
(1/2)\slash{p}\slash{z}\psi$ with 
spin projections $s=+1/2$ and $s=-1/2$, respectively. For the gluon 
field strength there are three possibilities:
 $G_{.\perp}$ corresponds to $s=+1$, 
$G_{*\perp}$ to $s=-1$ and both
$G_{\perp\perp}$ and $G_{.*}$ correspond to $s=0$.

Multi-particle states built of quasi-primary fields can be expanded in 
irreducible unitary representations with increasing conformal spin.
An explicit expression for the distribution amplitude 
of a multi-particle state with the lowest conformal spin 
 $j=j_1+\ldots+j_m$ built of $m$ primary fields with spins $j_k$ is
\begin{equation}
\phi_{as}(\alpha_1,\alpha_2,\ldots,\alpha_m) = 
\frac{\Gamma[2j_1+\ldots +2j_m]}{\Gamma[2j_1]\ldots \Gamma[2j_m]}
\alpha_1^{2j_1-1}\alpha_2^{2j_2-1}\ldots \alpha_m^{2j_m-1}.
\label{eq:asymptotic}
\end{equation}
Here $\alpha_k$ are the corresponding momentum fractions.
This state is nondegenerate and cannot mix with 
other states because of conformal symmetry.
Multi-particle irreducible representations with higher spin
$j+n,n=1,2,\ldots$, 
are given by  polynomials of $m$ variables (with the constraint 
$\sum_{k=1}^m \alpha_k=1$ ), which are orthogonal over
 the weight-function (\ref{eq:asymptotic}).

\section{Meson Mass Corrections}
\setcounter{equation}{0}

The structure of meson mass corrections in exclusive processes is 
in general more complicated than that of target mass corrections in 
deep inelastic scattering where they can be resummed using the
Nachtmann variable \cite{N73}. For illustration, consider the simplest 
matrix element
\begin{eqnarray}
\lefteqn{
 \langle0|\bar u(x)\xslash d(-x)|\rho^-(P,\lambda)\rangle =}\makebox[1cm]{\ }
\nonumber\\&=& f_\rho m_\rho
  (e^{(\lambda)}x)\int_0^1 \!du\, e^{i(2u-1)Px}
  \Big[\phi_\parallel(u)+\frac{x^2}{4}\Phi(u)+O(x^4)\Big].
\label{mass1}
\end{eqnarray}
We assume that $x^2 \ll \Lambda_{\rm QCD}^{-2}$, but nonzero, 
$\phi_\parallel(u)$ 
is the twist-2 chiral-even distribution amplitude and 
$\Phi(u)= m^2_\rho [{\Bbb A}(u)+ 
(1/2) \int_0^u\!dv\!\int_0^v\!dw\,{\Bbb C}(w)]$ (c.f. (\ref{eq:OPEvector})) 
describes
higher twist corrections, the ``kinematic'' contributions to which, due
to the massive $\rho$ meson, we want to calculate.

Experience with inclusive distributions 
tells us that meson mass corrections are related to contributions
of leading twist operators. Indeed, the conditions of symmetry and zero traces
for twist-2 local operators imply 
\begin{eqnarray}
\lefteqn{
\langle0|\Big[\bar u\xslash (i\deriv x)^n d\Big]_{\rm tw.2}
|\rho^-(P,\lambda)\rangle = }\makebox[1cm]{\ }
\nonumber\\&=&f_\rho m_\rho
(e^{(\lambda)}x)\Bigg[(Px)^n-\frac{x^2m_\rho^2}{4}\frac{n(n-1)}{n+1}(Px)^{n-2}
\Bigg]\langle\!\langle O_n\rangle\!\rangle,
\label{mass2}
\end{eqnarray} 
where $[\ldots]_{\rm tw.2}$ denotes taking the leading twist part (subtraction
of traces, in this case) and $\langle\!\langle O_n\rangle\!\rangle$ is
the reduced matrix element related to the $n$-th moment of the leading twist 
distribution 
\begin{equation}
  M_n^{(\parallel)}\equiv \int_0^1 \!du\,(2u-1)^n\phi_\parallel(u) = 
  \langle\!\langle O_n\rangle\!\rangle.
\end{equation}
Expanding (\ref{mass1}) at short distances $x\rightarrow0$ and
comparing it with 
(\ref{mass2}), we find that the same reduced matrix element gives a 
contribution to the twist-4 distribution amplitude:
\begin{equation}
  M_n^{(\Phi)}\equiv \int_0^1 \!du\,(2u-1)^n\Phi(u) = 
  \frac{1}{n+3}m_\rho^2\langle\!\langle O_{n+2}\rangle\!\rangle,
\label{mass3}
\end{equation}
which is the direct analogue of Nachtmann's correction in deep inelastic
scattering. 

As pointed out in~\cite{BB91}, there exists an alternative possibility 
to describe the mass corrections by modification of  the exponential
factor in (\ref{mass1}) rather than a contribution to the twist-4
distribution amplitude. To this end, we write
\begin{equation}
 \langle0|\Big[\bar u(x)\xslash d(-x)\Big]_{\rm tw.2}
         |\rho^-(P,\lambda)\rangle =
   f_\rho m_\rho
   \int_0^1 \!du\, \Big[(e^{(\lambda)}x) e^{i\xi Px}\Big]_{\rm tw.2}
   \phi_\parallel(u),
\label{fringe1}
\end{equation}
where $[\ldots]_{\rm tw.2}$ on the 
left-hand side correspond by definition to a subtraction of traces in all 
local operators in the  Taylor expansion of the nonlocal 
operator at short distances. As shown in~\cite{BB89}, this definition
implies that the nonlocal operator satisfies the homogeneous Laplace equation
\begin{equation}
  \frac{\partial^2}{\partial x_\eta\partial x^\eta} 
   \Big[\bar u(x)\xslash d(-x)\Big]_{\rm tw.2} = 0,
\end{equation} 
 and the same condition has to be fulfilled by the function 
$ \Big[(e^{(\lambda)}x) e^{i\xi Px}\Big]_{\rm tw.2}$ in order that 
Eq.~(\ref{fringe1}) be satisfied. The solution can easily be constructed 
order by order in the $(m^2x^2)^k$ expansion~\cite{BB91}. To
twist-4 accuracy, we obtain
\begin{equation}
  \Big[(e^{(\lambda)}x) e^{i\xi Px}\Big]_{\rm tw.2} =
     (e^{(\lambda)}x)\Big[ e^{i\xi Px} +\frac{m^2x^2\xi^2}{4}
    \int_0^1 \!dt\,t^2\,e^{it\xi Px} + O(x^4)\Big].
\label{exp_lt}
\end{equation}
By taking moments, it is easy to check that Eq.~(\ref{mass1}), with the 
higher twist distribution function $\Phi(u)$ given in (\ref{mass3}),
is equivalent to Eq.~(\ref{fringe1}) with the substitution (\ref{exp_lt}).  

The result in (\ref{mass3}) is, however, incomplete. The reason is that 
in exclusive processes one has to take into account higher twist operators 
containing total derivatives, and vacuum-to-meson matrix elements of 
such operators reduce, in certain cases, to powers of the meson mass times
reduced matrix elements of leading twist operators. 
In the present case, write \cite{BB89}
\begin{eqnarray}
  \bar u(x)\xslash d(-x) &=& \Big[\bar u(x)\xslash d(-x)\Big]_{\rm tw.2}+
\frac{x^2}{4}\int_0^1\!\!dt\,
\frac{\partial^2}{\partial x_\alpha \partial x^\alpha}\bar u(tx)\xslash d(-tx)
+O(x^4)
\nonumber\\
&=& \Big[\bar u(x)\xslash d(-x)\Big]_{\rm tw.2} -
\frac{x^2}{4}\int_0^1\!\!dt\,t^2\,\partial^2[\bar u(tx)\xslash d(-tx)]
\nonumber\\
&&+{\rm ~contributions~of~operators~with~gluons}+O(x^4),
\end{eqnarray} 
where we used Eq.~(\ref{twoderiv}) to obtain the last line.
In the matrix element we can make the substitution 
$\partial^2\rightarrow -m_\rho^2$.
Expanding, again, at short distances, and comparing with the short-distance
expansion of (\ref{mass1}), we get an additional 
contribution to $M_n^{(\Phi)}$, so that the corrected version of 
(\ref{mass3}) becomes 
\begin{equation}
  M_n^{(\Phi)} = 
  \frac{1}{n+3}m_\rho^2\,\big[\langle\!\langle O_{n+2}\rangle\!\rangle
    +\langle\!\langle O_{n}\rangle\!\rangle\big]+{\rm ~gluons}.
\label{mass4}
\end{equation}
Assuming the asymptotic form of the leading twist 
distribution amplitude $\phi$,
$\phi(u)=6u(1-u)$, so that  
$\langle\!\langle O_{n}\rangle\!\rangle =3/[2(n+1)(n+3)]$, this equation 
for moments is easily solved and gives
\begin{equation}
  \Phi(u) = 30 u^2(1-u)^2
  \left[\frac{2}{5}m_\rho^2 +\frac{4}{3}m^2_\rho\zeta_4\right],
\label{mass5}
\end{equation}
where we have included the ``genuine'' twist-4 correction (term in $\zeta_4$)
due to the twist-4 quark--gluon operator, see definition in 
Eq.~(\ref{def:zeta34}). 
The QCD sum rule estimate is $\zeta_4\sim 0.15$~\cite{BK86}, so that the meson 
mass effect on the twist-4 distribution function 
 is by a factor 2 larger than the ``genuine'' twist-4 correction. 
This is an important difference to deep inelastic scattering, where the 
target mass corrections are small.

The above discussion is still oversimplified and does not provide us with 
a complete separation of meson mass effects. The major complication 
arises because of  contributions of operators of the type
\begin{equation}
\partial_{\mu_1} \Big[\bar u\gamma_{\mu_1}(i\!\deriv_{\mu_2})
\ldots(i\!\deriv_{\mu_n})d\Big]_{\rm tw.2}.
\end{equation}
Such operators can be expressed in terms of operators with extra gluon fields,
which means, conversely, that certain combinations of $\bar qGq$ operators 
reduce to divergences of leading twist operators and give rise to extra 
meson mass correction terms. The corresponding corrections to twist-4 
distributions involve, however, 
 higher order contributions in the conformal expansion of the distribution 
amplitudes of leading twist and do not affect the result in (\ref{mass5}),
which only includes  leading conformal spin\footnote{
The reason why leading conformal spin is not affected is that the
 divergence of a conformal operator vanishes in free theory.}.
A calculation of the next-to-leading corrections will be presented below.

\section{Chiral-Even Distribution Amplitudes}
\setcounter{equation}{0}

In this section we derive explicit expressions for chiral-even
distribution amplitudes of twist-4 including the
leading and next-to-leading contributions in the conformal 
expansion. We first give a short summary of the relevant results of 
\cite{BBKT} to twist-3 accuracy. We then discuss the
conformal expansion of twist-4   three-particle distribution amplitudes
and relate the coefficients to
matrix elements of local operators. Only a few operators turn out to be 
independent, so that the number of nonperturbative parameters is reduced
considerably. Finally, we calculate the twist-4 two-particle
distribution amplitudes from the equations of motion (EOM).
The quark mass corrections will be neglected throughout this section.

\subsection{Twist-3 Distributions}

A comprehensive study of  $\rho$ meson distribution amplitudes 
to twist-3 accuracy was carried out in~\cite{CZreport,BBrho,BBKT},
and we begin this section by quoting the results relevant to the
present paper. 

The leading twist-2 distribution amplitude for the longitudinally 
polarized $\rho$ mesons, $\phi_\parallel$, is expanded as~\cite{CZreport,BBrho}
\begin{equation}\label{eq:phipar}
\phi_\parallel(u) =  6 u\bar u \left[ 1 +
a_2^\parallel\, \frac{3}{2} ( 5\xi^2  - 1 ) \right].
\end{equation}
The parameter $a_{2}^\parallel$ is defined by the  matrix
element of a twist-2 conformal operator with conformal spin 3:
\begin{eqnarray}
\langle 0|\bar u \slash{z} (i\deriv z)^2  d -\frac{1}{5}(i \partial z)^2
     \bar u \slash{z} d|\rho^-(P,\lambda)\rangle 
& = & \frac{12}{35}(e^{(\lambda)} z) (p z)^2 f_\rho m_\rho \,a_2^\parallel,
\label{eq:a12}
\end{eqnarray}
and is scale-dependent:
\begin{equation}
 a_2^\parallel(Q^2) = L^{\gamma^\parallel_2/b} a_2^\parallel(\mu^2),
~~~~
\gamma^\parallel_2 = \frac{25}{6}C_F,\label{eq:4.3}
\end{equation} 
where $L\equiv \alpha_s(Q^2)/\alpha_s(\mu^2)$ and $C_F=(N_c^2-1)/(2N_c)$,
$b=(11 N_c-2 n_f)/3$. The parameter
$a_2^\parallel$ has been estimated from QCD sum
rules; its value at the reference scale $\mu=1\,$GeV is given in
Table~\ref{tab:para1}. 

The three-particle distributions of twist-3 read~\cite{CZreport,BBKT}\footnote{
Note that we use a  normalization of distribution amplitudes 
different from that in \cite{CZreport,BBKT}. 
In the notation of Ref.~\cite{BBKT}, $\omega_{1,0}^A\equiv \omega_3^A$, 
$ \zeta_3^A\equiv \zeta_3$, and 
$\zeta_3^V \equiv (3/28)\zeta_3\omega_3^V$.}:
\begin{eqnarray}
{\cal V} (\underline{\alpha}) &=& 
540\, \zeta_3 \omega^V_3 (\alpha_d-\alpha_u)\alpha_d \alpha_u\alpha_g^2,
\label{modelV}\\
{\cal A} (\underline{\alpha}) &=& 
360\,\zeta_3 \alpha_d \alpha_u \alpha_g^2 
\left[ 1+ \omega^A_{3}\,\frac{1}{2}\,(7\alpha_g-3)\right].
\label{modelA}
 \end{eqnarray}
The dimensionless coupling $\zeta_3$ is
defined by the  matrix element
\begin{eqnarray}
\langle0|\bar u g\tilde G_{\mu\nu}\gamma_\alpha
 \gamma_5 d|\rho^-(P,\lambda)\rangle & = & 
f_\rho m_\rho \zeta_{3}
\Bigg[
e^{(\lambda)}_\mu\Big(P_\alpha P_\nu-\frac{1}{3}m^2_\rho \,g_{\alpha\nu}\Big)
-e^{(\lambda)}_\nu\Big(P_\alpha P_\mu-\frac{1}{3}m^2_\rho \,g_{\alpha\mu}\Big)
\Bigg]\nonumber\\
& & {}+ \frac{1}{3}f_\rho m_\rho^3 \zeta_{4}
\Bigg[e^{(\lambda)}_\mu g_{\alpha\nu}- e^{(\lambda)}_\nu g_{\alpha\mu}\Bigg],
\label{def:zeta34}
\end{eqnarray}
where $\zeta_4$ is a matrix element of twist-4, which we will need
below, while
$\omega_3^V$ and $\omega_3^A$ are defined as
\begin{equation}
\langle 0|\bar u \slash{z} (gG_{\alpha\nu}
z^\alpha (i\derright z) - (i\derleft z) g G_{\alpha\nu}z^\alpha)
d |\rho^-(P,\lambda)\rangle  
 =  i(pz)^3 e^{(\lambda)}_{\perp\nu}
m_\rho f_\rho \frac{3}{28}\, \zeta_3 \,\omega_3^V + O(z_\nu)
\label{eq:4.7}\end{equation}
and 
\begin{equation}
\langle 0|\bar u \slash{z}\gamma_5 \left[ i Dz,g\tilde{G}_{\mu\nu}
  z^\mu \right] d -\frac{3}{7} (i\partial z)
\bar u \slash{z}\gamma_5 g\tilde{G}_{\mu\nu} z^\mu d
|\rho^-(P,\lambda)\rangle = -(pz)^3 e^{(\lambda)}_{\perp\nu} 
m_\rho f_\rho \frac{3}{28} \zeta_3\,\omega_3^A + O(z_\nu),
\label{eq:4.8}\end{equation}
respectively, where $[~,~]$ stands for the commutator.

The scale-dependence of the twist-3 parameters is given by~\cite{BBKT}
(with $C_A = N_c$):
\begin{equation}
  \zeta_3(Q^2) = L^{\gamma_3^\zeta/b}\zeta_3(\mu^2),
~~~\gamma_3^\zeta = -\frac{1}{3}C_F +3 C_A,
\end{equation} 
and
\beq
& & \left(
\begin{array}{c}
\omega_{3}^{V} - \omega^{A}_3\\
\omega_{3}^{V} + \omega^{A}_3
\end{array}
\right)^{Q^{2}} =  L^{\Gamma_3^\omega/b}
\left(
\begin{array}{c}
\omega_{3}^{V} - \omega^{A}_3\\
\omega_{3}^{V} + \omega^{A}_3
\end{array}
\right)^{\mu^{2}},\nonumber\\
& &\Gamma_3^\omega = \left(
\begin{array}{cc}
 3C_{F} - {2\over 3}C_{A}\ \ &
\frac{2}{3}C_{F}-\frac{2}{3}C_{A} \\
\frac{5}{3}C_{F} - \frac{4}{3}C_{A}\ \ &
\frac{1}{2}C_{F} + C_{A}
\end{array}
\right).
\label{eq:4exam2}
\eeq
Numerical estimates are given in Table~\ref{tab:para1}.

Finally, the two-particle distributions of twist-3 are determined from
the EOM \cite{BBKT}:
\begin{eqnarray}
g_\perp^{(a)}(u) & = & 6 u \bar u \!\left[ 1 + 
\!\left\{\frac{1}{4}a_2^\parallel +
\frac{5}{3}\, \zeta_{3} \!\left(1-\frac{3}{16}\,
\omega^A_{3}+\frac{9}{16}\omega^V_3\!\right)\!\right\}
(5\xi^2-1)\right],\nonumber\\
 g_\perp^{(v)}(u) & = & \frac{3}{4}(1+\xi^2)
+ \left(\frac{3}{7} \, 
a_2^\parallel + 5 \zeta_{3} \right) \left(3\xi^2-1\right)
 \nonumber\\
& & {}+ \left[ \frac{9}{112}\, a_2^\parallel 
+ \frac{15}{64}\, \zeta_{3}\Big(3\,\omega_{3}^V-\omega_{3}^A\Big)
 \right] \left( 3 - 30 \xi^2 + 35\xi^4\right).
\label{eq:gv}
\end{eqnarray}

\subsection{Twist-4 Distributions}

Due to the odd G-parity of the operator in (\ref{eq:even2}), the
distribution amplitudes  $\Phi$ and $\Psi$ are antisymmetric under the
exchange of $\alpha_d$ and $\alpha_u$, whereas $\widetilde{\Phi}$ and  
$\widetilde{\Psi}$ are symmetric.
The distributions $\widetilde{\Psi}$, $\Psi$ correspond to
the light-cone projection $\gamma_\cdot G_{\cdot *}$ (see Table~1) and
have the conformal expansion
\begin{eqnarray}
\widetilde{\Psi}(\underline{\alpha}) &=& 120\alpha_u\alpha_d\alpha_g
\left[ \widetilde{\psi}_{00} + \widetilde{\psi}_{10} (3\alpha_g-1)
+\ldots\right],
\nonumber\\
\Psi(\underline{\alpha})& =& 120 \alpha_u\alpha_d\alpha_g\,  
\Big[ ~0~+\,\psi_{10}
(\alpha_d-\alpha_u) + \ldots\Big]\,,\label{eq:Psi}
\end{eqnarray}
respectively. Note that the leading spin contribution to $\Psi$ vanishes 
because of G-parity (for massless quarks).

In turn, the distribution amplitudes $\widetilde{\Phi}$,  $\Phi$
correspond to the $\gamma_\perp G_{\cdot\perp}$ light-cone component, 
and before a conformal expansion  can be performed,  we  first
have to separate the 
different quark spin projections. To this end, we define auxiliary amplitudes:
\begin{eqnarray}
\langle 0 | \bar u(z) g\widetilde{G}_{\mu\nu}(vz) \gamma_\cdot
\gamma_\alpha\gamma_5\gamma_* d(-z) | \rho\rangle & = & f_\rho
m_\rho^3 \,\frac{ez}{pz}\left( p_\mu g^\perp_{\alpha\nu} - p_\nu
  g^\perp_{\alpha\mu} \right)
\Phi^{\uparrow\downarrow}(v,pz),\nonumber\\
\langle 0 | \bar u(z) g\widetilde{G}_{\mu\nu}(vz) \gamma_*
\gamma_\alpha\gamma_5\gamma_\cdot d(-z) | \rho\rangle & = & f_\rho
m_\rho^3 \,\frac{ez}{pz}\left( p_\mu g^\perp_{\alpha\nu} - p_\nu
  g^\perp_{\alpha\mu} \right)
\Phi^{\downarrow\uparrow}(v,pz).
\end{eqnarray}
The relation of $\Phi^{\ud},\Phi^{\du}$ to the original amplitudes
is given by:
\begin{eqnarray}
\widetilde{\Phi}(\underline{\alpha}) & = &
\frac{1}{2}\left[\Phi^{\uparrow\downarrow} +
  \Phi^{\downarrow\uparrow}\right](\underline{\alpha}),\nonumber\\
\Phi(\underline{\alpha}) & = &
\frac{1}{2}\left[\Phi^{\uparrow\downarrow} -
  \Phi^{\downarrow\uparrow}\right](\underline{\alpha}),
\end{eqnarray}
and their conformal expansion goes in terms of Appell polynomials:
\begin{eqnarray}
\Phi^{\uparrow\downarrow}(\underline{\alpha}) & = &
60\alpha_u\alpha_g^2 \left[ \phi_{00} + \phi_{01}(\alpha_g-3\alpha_d) +
  \phi_{10}\left(\alpha_g-\frac{3}{2}\alpha_u\right)\right],\nonumber\\
\Phi^{\downarrow\uparrow}(\underline{\alpha}) & = &
60\alpha_d\alpha_g^2 \left[ \phi_{00} +\phi_{01}(\alpha_g-3\alpha_u) +
  \phi_{10}\left(\alpha_g-\frac{3}{2}\alpha_d\right)\right],\label{eq:Phi}
\end{eqnarray}
where we have taken into account the symmetry properties, i.e.\ 
$\phi_{00}^{\ud} = \phi_{00}^{\du}$, etc. Combining everything, 
we obtain
\begin{eqnarray}
\widetilde{\Phi}(\underline{\alpha}) & = &
 30 \alpha_g^2\left[ \phi_{00}(1-\alpha_g)
                    +\phi_{01}\Big[\alpha_g(1-\alpha_g)-6\alpha_u\alpha_d\Big]
                    +\phi_{10}\Big[\alpha_g(1-\alpha_g)-\frac{3}{2}(\alpha_u^2
                               +\alpha_d^2)\Big]\right],
\nonumber\\
\Phi(\underline{\alpha}) & = &
 30 \alpha_g^2(\alpha_u-\alpha_d)\left[ \phi_{00}
                    +\phi_{01}\alpha_g
                    +\frac{1}{2}\phi_{10}(5\alpha_g-3)
                                 \right].
\end{eqnarray}

At this point, the expansion involves two parameters, $\phi_{00}$ and
$\wt{\psi}_{00}$, to leading conformal twist accuracy, and 
four more ($\phi_{10},\phi_{01},\psi_{10},\widetilde\psi_{10}$)
for the corrections. Our next task will be to relate them 
 to matrix elements of local operators and 
find out how many coefficients are independent.

For leading spin the answer is easily obtained by taking 
the relevant light-cone projections of the matrix element 
in (\ref{def:zeta34}):
\begin{eqnarray}
  \widetilde{\psi}_{00} &=&  \phantom{-}\frac{2}{3}\,\zeta_3 + \frac{1}{3}\,
   \zeta_4,
\nonumber\\
   \phi_{00} &=&  -\frac{1}{3}\,\zeta_3 + \frac{1}{3}\, \zeta_4.
\end{eqnarray}
Note that the ``twist-4'' distribution
amplitudes receive contributions of {\em both} twist-3 and twist-4
operators. This is due to the fact that the standard counting of twist
in terms of ``good'' and ``bad'' components as introduced in \cite{KS}
is at variance with the definition of twist as spin minus dimension of
an operator. See also the discussion in Sec.~2.2.\ of
Ref.~\cite{BBKT}. The parameter $\zeta_4$ in 
(\ref{def:zeta34}) can be explicitly 
defined as the matrix element of a pure twist-4 operator:
\begin{equation}
\langle0|\bar u g\tilde G_{\mu\nu}\gamma_\nu
 \gamma_5 d|\rho^-(P,\lambda)\rangle =
   f_\rho m_\rho^3 e^{(\lambda)}_\mu \zeta_{4}.
\label{def:zeta4}
\end{equation}
Its scale-dependence is given by~\cite{SV82}
\begin{equation}
  \zeta_4(Q^2) = L^{\gamma_4^\zeta/b}\zeta_4(\mu^2),
~~~\gamma_4^\zeta = \frac{8}{3}C_F,
\end{equation}
and the numerical value was estimated in~\cite{BK86} from QCD sum
rules, see Table~\ref{tab:para1} and App.~C. 

The calculation of the next-to-leading order spin 
corrections is more involved and is presented 
in detail in App.~B. The main observation is that the four coefficients 
$\phi_{10},\phi_{01},\psi_{10},\widetilde\psi_{10}$ in fact involve 
only one new nonperturbative parameter, in addition to the ones defined
above. We find:
\begin{eqnarray}
 \phi_{01}&=& \phantom{-}\frac{1}{12}a_2^\parallel -\frac{5}{12}\zeta_3
              +\frac{3}{16}\,\zeta_3(\omega_3^A+\omega_3^V)
              + \frac{7}{2}{\zeta_4\, \omega_4^A},
\nonumber\\
 \phi_{10}&=& -\frac{1}{12}a_2^\parallel +\frac{3}{4}\,\zeta_3
              +\frac{3}{16}\zeta_3(\omega_3^A-\omega_3^V)
              + 7 {\zeta_4\, \omega_4^A},
\nonumber\\
 \psi_{10}&=& -\frac{1}{4}a_2^\parallel -\frac{7}{12}\,\zeta_3
              +\frac{9}{16}\zeta_3\,\omega_3^V
              - \frac{21}{4}{\zeta_4\, \omega_4^A},
\nonumber\\
 \widetilde\psi_{10}&=& \phantom{-} \frac{2}{3}\,\zeta_3
              -\frac{9}{16}\zeta_3\,\omega_3^A
              + \frac{21}{4}{\zeta_4\, \omega_4^A},
\label{eq:NLO}
\end{eqnarray}
where the new parameter $\omega_4^A$ is defined as
\begin{eqnarray}
 \lefteqn{\langle0|
 \bar u \left [iD_\mu, g\tilde G_{\nu\xi}\right]\gamma_\xi \gamma_5 d 
-\frac{4}{9}(i\partial_\mu)
\bar u g\tilde G_{\nu\xi}\gamma_\xi \gamma_5 d 
|\rho^-(P,\lambda)\rangle +(\mu\leftrightarrow\nu) =}
\nonumber\\
&&{}\hspace*{6cm}= 
2 f_\rho m_\rho^3 \,\zeta_4\, \omega_4^A
 \left(e^{(\lambda)}_\mu P_\nu + e^{(\lambda)}_\nu P_\mu \right).
\mbox{\hspace{0cm}}\label{eq:w4A}
\end{eqnarray}
$\omega_4^A$ is estimated from QCD sum rules in App.~C, with the result
given in Table~\ref{tab:para1}. The one-loop anomalous dimension of
the operator on the left-hand side of (\ref{eq:w4A}) is not known.

\begin{table}
\renewcommand{\arraystretch}{1.4}
\addtolength{\arraycolsep}{3pt}
$$
\begin{array}{|ccccccc|}\hline
f_\rho\,[{\rm MeV}] & a_2^\parallel & \zeta_3 & \omega_3^A &
\omega_3^V & \zeta_4 & \omega_4^A\\ \hline
198\pm 7 & 0.18\pm 0.10 & 0.032\pm 0.010 &
-2.1\pm1.0 & 3.8\pm1.8 & 0.15\pm 0.10 & 0.8\pm 0.8\\ \hline
\end{array}
$$
\caption[]{Parameters of chiral-even distribution amplitudes.
  Renormalization scale is $\mu = 1\,$GeV.}\label{tab:para1}
\renewcommand{\arraystretch}{1}
\addtolength{\arraycolsep}{-3pt}
\end{table}

A few comments on the structure of (\ref{eq:NLO}) are 
in order. 
First, as already mentioned, twist-4 distribution amplitudes contain 
contributions of operators of twist-3. 
Note that the twist-4 chiral-even distributions considered here correspond
to longitudinally polarized $\rho$ mesons, while the twist-3
parts appearing in (\ref{eq:NLO}) formally correspond to transversely 
polarized mesons. The physical reason why  distributions with 
different polarization appear to be  related is 
 Lorentz symmetry: a longitudinally polarized $\rho$ meson can 
be made transversely polarized by going over to the meson rest frame,
rotating the spin and boosting back to the infinite momentum 
frame. The spin rotation, however, is not a member of the collinear 
conformal group. Because of this, the conformal structure of twist-3
additions to twist-4 amplitudes is rather complicated and does not 
match the naive expansion, similar to the case considered 
in App.~B of \cite{BBKT}. Formally, this is yet another complication 
of having a nonzero meson mass.

Secondly, we find a term in $a_2^\parallel$ that 
corresponds to the next-to-leading correction in the 
conformal expansion of the leading twist distribution amplitude.
This contribution thus presents an additional meson mass correction 
and appears, in technical terms, through the operator
identity (see App.~B) relating the divergence of a two-particle 
conformal operator to operators involving gluon fields.
In this respect
distribution amplitudes in exclusive reactions are fundamentally
different from inclusive distributions, which involve only
forward-scattering matrix elements, so that matrix elements of 
operators with total derivatives vanish. 

Third,  an inspection of the numerical size of the 
entries in Table~\ref{tab:para1} reveals 
that the coefficients in (\ref{eq:NLO}) are 
grossly dominated by the  term in $\omega_A^4$, which is 
a genuine twist-4 effect. We do not see any physical reasons for 
this dominance, but, if correct, it suggests that the above-mentioned
complications may have an only marginal effect on phenomenology. 

Finally, we have to specify the two-particle twist-4 distributions 
$g_3$ and $\Bbb A$ defined 
in Sec.~2. They are not independent, but can be expressed 
in terms of  $\Phi$ and $\Psi$ by using the EOM, see Eqs.~(\ref{eq:rel1even}),
(\ref{eq:rel2even}). To next-to-leading accuracy, we obtain:
\begin{eqnarray}
g_3(u) & = & 1 + \left( -1-\frac{2}{7}\,a_2^\parallel + \frac{40}{3}\,
  \zeta_3 - \frac{20}{3}\,\zeta_4\right) C_2^{1/2}(\xi)\nonumber\\
&& + \left
  ( -\frac{27}{28} \, a_2^\parallel + \frac{5}{4}\,\zeta_3 -
  \frac{15}{16}\, \zeta_3 \left\{ \omega_3^A + 3\omega_3^V\right\}
  \right) C_4^{1/2}(\xi),\label{eq:expg3}\\
{\Bbb A}(u) & = & 30 u^2\bar u^2 \left\{\frac{4}{5}\left(
   1+\frac{1}{21}\, a_2^\parallel + \frac{10}{9}\,\zeta_3 +
  \frac{25}{9}\,\zeta_4\right)\right.\nonumber\\
&&\left.{} + \frac{1}{5}
\left( \frac{9}{14}\, a_2^\parallel +
  \frac{1}{18}\,\zeta_3 + \frac{3}{8}\,\zeta_3 \left[\frac{7}{3}\,
  \omega_3^V - \omega_3^A\right]\right)
  C_2^{5/2}(\xi)\right\}\nonumber\\
 & & {} + 10
  \left(-2a_2^\parallel - \frac{14}{3} \,\zeta_3 +
  \frac{9}{2}\,\zeta_3\omega_3^V - 42\zeta_4
  \omega_4^A\right)\nonumber\\
  & & \times
  \int_0^u dv \int_0^v dw \, \left\{ 1 + C_2^{1/2}(\xi_w) - 3 \xi_w
  (1-\xi_w^2) \ln \,\frac{1+\xi_w}{1-\xi_w}\right\}\label{eq:expA}
\end{eqnarray}
with $\xi_w = 2w-1$.
The double-integral can of course be taken analytically: 
\begin{eqnarray}
\lefteqn{\hspace*{-1cm}
\int_0^u dv \int_0^v dw \, \left\{ 1 + C_2^{1/2}(\xi_w) - 3 \xi_w
  (1-\xi_w^2) \ln \,\frac{1+\xi_w}{1-\xi_w}\right\} =}
\nonumber\\
&=& \frac{1}{10}u\bar u(2+13u\bar u)+
   \frac{1}{5} u^3(10-15 u+6 u^2)\ln  u +
   \frac{1}{5}\bar u^3(10-15\bar u+6\bar u^2)\ln \bar u.
\end{eqnarray}
The resulting functions $g_3(u)$ and ${\Bbb A}(u)$ are shown in Fig.~1
by solid lines. 
The dashed  curves are obtained by omitting the next-to-leading 
spin corrections (which is the approximation adopted in  
\cite{handbook,BB98}), and the dash-dotted curves correspond to 
taking into account the 
meson mass corrections only and neglecting all twist-3 and twist-4 
matrix elements.
It is evident that the contributions from next-to-leading
order conformal spin are small in both
cases. The mass terms clearly dominate $g_3(u)$ and constitute
about half of ${\Bbb A}(u)$.
\begin{figure}
\centerline{\epsffile{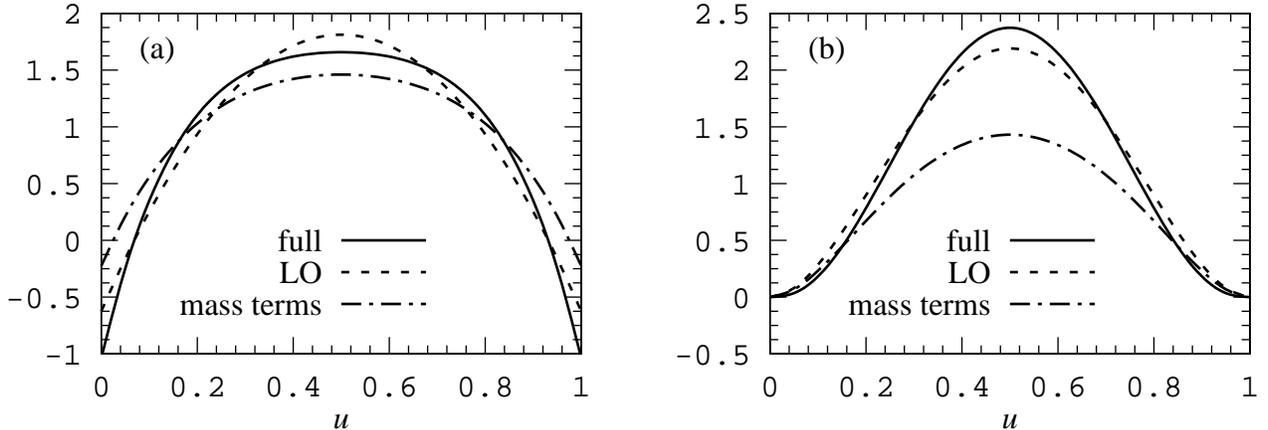}}
\caption[]{
Two-particle twist-4 chiral-even distribution amplitudes of the $\rho$
meson: $g_3$ (a) and $\Bbb A$ (b). LO means neglecting contributions
of higher conformal spin for twist-3 and twist-4 operators and 
the mass terms correspond to pure meson mass corrections.
}
\end{figure}

We stress that the given expressions are exact, provided the three-particle 
distributions are taken in the above approximation. This means, 
in particular, that (\ref{eq:expg3}) and (\ref{eq:expA}) reproduce the 
exact second moments of $g_3$ and $d^2/du^2 {\Bbb A}$, i.e.\ the
normalization of $\Bbb A$, but the fourth moment of $g_3$
(second of $\Bbb A$) also includes (uncalculated) contributions 
from even higher conformal spin operators. We have checked that the
second moments agree with those obtained from Taylor expanding
(\ref{eq:OPEvector}). 

Note that $g_3$ corresponds to the
spin projection $s=-1/2$ for both the quark and the antiquark, 
and thus has a
conformal expansion in Gegenbauer polynomials $C^{1/2}(2u-1)$, 
cf.\ (\ref{eq:asymptotic}):
$$
g_3(u,\mu^2) = 1 + \sum_{k=2,4,\ldots}^\infty g^{(k)}_3(\mu^2)
C^{1/2}_{k}(2u-1).
$$ 
The coefficients $g^{(2)}_3$ and $g^{(4)}_3$ can be read off
(\ref{eq:expg3}). The conformal expansion of ${\Bbb A}$ is not 
straightforward and contains for instance logarithms. 

\section{Chiral-Odd Distribution Amplitudes}
\setcounter{equation}{0}

The construction of twist-4 chiral-odd distribution amplitudes parallels
that for chiral-even distributions  in the previous
section. We first recall the results for  distribution
amplitudes of twist-3. Next, we 
derive the conformal expansion of three-particle
distribution amplitudes to next-to-leading order in conformal spin and
relate the expansion coefficients to matrix elements of local
operators. The two-particle distribution amplitudes are then
obtained from the EOM as derived in App.~A.

\subsection{Twist-3 Distributions}

The leading twist-2 distribution
amplitude for  transversely
polarized $\rho$ mesons, $\phi_\perp$, is expanded as~\cite{CZreport,BBrho}
\begin{equation}\label{eq:phiperp}
\phi_\perp(u) =  6 u\bar u \left[ 1 +
a_2^\perp\, \frac{3}{2} ( 5\xi^2  - 1 ) \right].
\end{equation}
The parameter $a_{2}^\perp$ is defined by the  matrix
element of a twist-2 conformal operator with conformal spin 3,
\begin{eqnarray}
\langle 0|\bar u \sigma_{\perp\cdot} (i\deriv z)^2  d -
\frac{1}{5}(i \partial z)^2 \bar u \sigma_{\perp\cdot}
 d|\rho^-(P,\lambda)\rangle
& = & \frac{12}{35}\, e^{(\lambda)}_\perp 
(p z) i f_\rho^T  \,a_2^\perp,
\end{eqnarray}
and is scale-dependent:
\begin{equation}
 a_2^\perp(Q^2) = L^{\gamma^\perp_2/b} a_2^\perp(\mu^2),
~~~~
\gamma^\perp_2 = \frac{10}{3}C_F.
\end{equation}
 The numerical value of $a_2^\perp$ has been estimated from QCD sum
rules and is given in
Table~\ref{tab:para2} (at the reference scale $\mu=1\,$GeV). 

The only existing three-particle distribution amplitude of twist-3, 
${\cal T}$, is given by:
\begin{equation}
{\cal T}(\underline{\alpha}) = 540 \,\zeta_3\, \omega_3^T
(\alpha_d-\alpha_u) \alpha_d \alpha_u \alpha_g^2,
\end{equation}
with $\omega_3^T$ defined as 
\begin{equation}
\langle 0|\bar u \sigma_{\mu\nu}z^\nu (gG^{\mu\beta}
z_\beta (i\!\derright\! z) - (i\!\derleft\! z) g G^{\mu\beta}z_\beta)
d |\rho^-(P,\lambda)\rangle  = (pz)^2 (e^{(\lambda)}z)
m_\rho^2 f_\rho^T \frac{3}{28}\, \zeta_3 \omega_3^T.
\end{equation}
The parameter $\zeta_3$ was already defined in (\ref{def:zeta34}). The
scale-dependence of $\omega_3^T$ is given by \cite{BBKT}
\begin{equation}
\omega_3^T(Q^2) = L^{\gamma_3^{\omega^T}/b} \omega_3^T(\mu^2), \quad
\gamma_3^{\omega^T} = \frac{25}{6} \, C_F - 2 C_A.
\end{equation}
Finally, we also quote the two-particle distributions of twist-3 as
obtained from the EOM \cite{BBKT}:
\begin{eqnarray}
h_\parallel^{(s)}(u) & = & 6u\bar u \left[ 1 + \left( \frac{1}{4}a_2^\perp +
\frac{5}{8}\,\zeta_{3}\omega_3^T \right) (5\xi^2-1)\right],\\
h_\parallel^{(t)}(u) &= & 3\xi^2+ 
 \frac{3}{2} a_2^\perp\, \xi^2 \,(5\xi^2-3)
 +\frac{15}{16}\zeta_{3}\omega_3^T(3-30\xi^2+35\xi^4).
\end{eqnarray}
Numerical values of the 
input parameters are collected in Table~\ref{tab:para2}.

\subsection{Twist-4 Distributions}

Due to the odd G-parity of the operator in (\ref{eq:T3}), the
distribution amplitudes $T_i^{(4)}$ are antisymmetric under the
exchange of $\alpha_u$ and $\alpha_d$, whereas $S$ and $\wt S$ are
symmetric. In order to resolve the conformal structure of $T_i^{(4)}$,
it is advantageous to exploit the fact that
$\sigma_{\mu\nu}\gamma_5$ is not independent of $\sigma_{\mu\nu}$, and
to define a ``dual'' matrix element 
\begin{equation}
\langle 0 | \bar u(z) i \sigma_{\alpha\beta} \gamma_5
g\wt{G}_{\mu\nu}(vz) d(-z) | \rho\rangle = \mbox{r.h.s.\ of
  (\ref{eq:T3}) with $T\to\wt{T}$}.
\end{equation}
One easily finds
\begin{eqnarray}
\wt{\cal T}^{(3)} & = & {\cal T}^{(3)},\quad \phantom{-}\wt{T}_1^{(4)}\ =\
-T_3^{(4)}, \quad \wt{T}_2^{(4)}\ =\ -T_4^{(4)},\nonumber\\
\wt{T}_3^{(4)} & = & -T_1^{(4)},\quad \wt{T}_4^{(4)}\ =\ -T_2^{(4)}.
\end{eqnarray}
We next note that 
the distributions $T_1^{(4)}$ and $\wt{T}_1^{(4)} = - T_3^{(4)}$
correspond to the Lorentz spin projection $s=+1/2$ for both quark
fields and the spin projection zero for the gluon. Hence
\begin{eqnarray}
T_1^{(4)}(\underline{\alpha}) & = & \phantom{-}120 t_{10} (\alpha_u-\alpha_d)
\alpha_u\alpha_d\alpha_g,\nonumber\\
T_3^{(4)}(\underline{\alpha}) & = & -120 \wt t_{10} (\alpha_u-\alpha_d)
\alpha_u\alpha_d\alpha_g.
\end{eqnarray}

For the distribution amplitudes $S$, $\wt S$, $T_1^{(4)}$ and $T_4^{(4)}$,
on the other hand, one has to separate different quark spin projections.
To this end, we define auxiliary amplitudes
\begin{eqnarray}
\langle 0 | \bar u(z) \gamma_\cdot \gamma_* gG_{\mu\nu}(vz)
d(-z)|\rho^-(P,\lambda)\rangle & = & if_\rho^T m_\rho^2 
[e^{(\lambda)\perp}_\mu p_\nu -
e^{(\lambda)\perp}_\nu p_\mu] S^{\ud}(v,pz),\nonumber\\
\langle 0 | \bar u(z) \gamma_\cdot \gamma_* i\gamma_5 g
\widetilde{G}_{\mu\nu}(vz)
d(-z)|\rho^-(P,\lambda)\rangle & = & if_\rho^T m_\rho^2 
[e^{(\lambda)\perp}_\mu p_\nu -e^{(\lambda)\perp}_\nu p_\mu] 
\wt{S}^{\ud}(v,pz),
\end{eqnarray}
and, similarly, two more distributions $S^{\du}$ and $\wt{S}^{\du}$ 
by replacing $\gamma_\cdot\gamma_*\rightarrow \gamma_*\gamma_\cdot$. 
The relations to the distribution amplitudes in (\ref{eq:T3}),
  (\ref{eq:2.21}) are given by:
\begin{eqnarray}
{ S}(\underline{\alpha}) & = & \frac{1}{2}\, ({
  S}^{\ud}(\underline{\alpha})  +
{ S}^{\du}(\underline{\alpha})),\nonumber\\
\wt{ S}(\underline{\alpha}) & = & \frac{1}{2}\, (\wt{
  S}^{\ud}(\underline{\alpha})  +
\wt{ S}^{\du}(\underline{\alpha})),\nonumber\\
T_4^{(4)}(\underline{\alpha}) & = & \frac{1}{2}\, ({
  S}^{\ud}(\underline{\alpha})  -
{ S}^{\du}(\underline{\alpha})),\nonumber\\
-T_2^{(4)}(\underline{\alpha})\ =\ \wt{T}_4^{(4)}(\underline{\alpha}) 
& = & \frac{1}{2}\, (\wt{ S}^{\ud}(\underline{\alpha})  -
\wt{ S}^{\du}(\underline{\alpha})).\label{eq:obvious}
\end{eqnarray}
The auxiliary amplitudes are expanded in Appell polynomials as
\begin{eqnarray}
{ S}^{\ud}(\underline{\alpha}) & = & 60 \alpha_u\alpha_g^2 \left
  [ s_{00} + s_{10} \left(\alpha_g-\frac{3}{2}\, \alpha_u\right) +
  s_{01} (\alpha_g - 3\alpha_d)\right],\nonumber\\
{ S}^{\du}(\underline{\alpha}) & = & 60 \alpha_d\alpha_g^2 \left
  [ s_{00} + s_{10} \left(\alpha_g-\frac{3}{2}\, \alpha_d\right) +
  s_{01} (\alpha_g - 3\alpha_u)\right],\label{eq:5.14}
\end{eqnarray}
and similarly for $\wt{ S}^{\ud}$ and $\wt{ S}^{\du}$. Here we
made use of the symmetry of $S$ under the exchange
 of the $u$ and $d$ quarks, i.e.\ $s_{00}^{\ud}
= s_{00}^{\du}$, etc.

{}From (\ref{eq:obvious}) and (\ref{eq:5.14}) it now follows 
immediately that
\begin{eqnarray}
{ S}(\underline{\alpha}) & = & 30\alpha_g^2\! \left
  [ s_{00}\,(1-\alpha_g) +
  s_{10} \left\{\!\alpha_g (1-\alpha_g) - \frac{3}{2}\, (\alpha_u^2 +
  \alpha_d^2) \right\} + s_{01} \left\{ \alpha_g (1-\alpha_g) -
  6\alpha_u\alpha_d\right\} \right],\nonumber\\
\wt{ S}(\underline{\alpha}) & = & 30\alpha_g^2\! \left[ \wt
  s_{00}\,(1-\alpha_g)  +
  \wt s_{10} \left\{\!\alpha_g (1-\alpha_g) - \frac{3}{2}\, (\alpha_u^2 +
  \alpha_d^2) \right\} + \wt s_{01} \left\{ \alpha_g (1-\alpha_g) -
  6\alpha_u\alpha_d\right\} \right],\nonumber\\
T_2^{(4)}(\underline{\alpha}) & = & -30\alpha_g^2 (\alpha_u-\alpha_d) 
\left[ \wt
  s_{00} + \frac{1}{2}\,\wt s_{10}\, (5\alpha_g-3) + \wt
  s_{01} \alpha_g\right],\nonumber\\
T_4^{(4)}(\underline{\alpha}) & = & \phantom{-}
30\alpha_g^2 (\alpha_u-\alpha_d) \left[ 
  s_{00} + \frac{1}{2}\, s_{10}\, (5\alpha_g-3) + 
  s_{01} \alpha_g\right].
\end{eqnarray}

At this point, the expansion involves two parameters of leading
conformal spin, $s_{00}$ and $\wt s_{00}$, and six more
($s_{10},s_{01},\wt s_{10},\wt s_{01},t_{10},\wt t_{10}$) for the
corrections. Our next task will be to relate them to matrix elements
of local operators and find out how many coefficients are independent.

For the leading spin, the answer is easily obtained by taking the
local limit $z\to 0$ of (\ref{eq:2.21}), so that
\begin{equation}
s_{00} = \zeta_4^T,\quad \wt s_{00} = \wt \zeta_4^T
\end{equation}
with
\begin{eqnarray}
\langle 0|\bar u gG_{\mu \nu}d|\rho^-(P,\lambda)\rangle &=&
  if_\rho^T m_\rho^2 \zeta^T_4(
e^{(\lambda)}_{\mu}P_\nu - e^{(\lambda)}_{\nu}P_\mu),
\nonumber\\
\langle 0|\bar u g\widetilde G_{\mu \nu}i\gamma_5
   d|\rho^-(P,\lambda)\rangle &=&
  if_\rho^T m_\rho^2 \widetilde \zeta^T_4(
e^{(\lambda)}_{\mu}P_\nu - e^{(\lambda)}_{\nu}P_\mu).\label{eq:defzetaT}
\end{eqnarray}
The parameters $\zeta_4^T$, $\wt\zeta_4^T$
renormalize multiplicatively \cite{BBK}:
\begin{eqnarray}
\left(\zeta_4^T + \wt{\zeta}_4^T\right)(Q^2) = L^{\gamma^+/b}
\left(\zeta_4^T + \wt{\zeta}_4^T\right)(\mu^2),& & \gamma_+ = 3 C_A -
\frac{8}{3}\, C_F,\nonumber\\
\left(\zeta_4^T - \wt{\zeta}_4^T\right)(Q^2) = L^{\gamma^-/b}
\left(\zeta_4^T - \wt{\zeta}_4^T\right)(\mu^2),& & \gamma_- = 4 C_A -
4 C_F.
\end{eqnarray}
The numerical values can be estimated from QCD sum rules, see
Table~\ref{tab:para3} and App.~C.

The calculation of the next-to-leading order spin corrections is
involved and presented in detail in App.~B. The main observation is
that the six coefficients $s_{10}$, $s_{01}$, $\wt s_{10}$, $\wt s_{01}$,
$t_{10}$ and $\wt t_{10}$ involve three new nonperturbative
parameters. We find:
\begin{eqnarray}
s_{10} & = & -\frac{3}{22}\, a_2^\perp - \frac{1}{8}\, \zeta_3
\omega_3^T + \frac{28}{55}\, \langle\!\langle Q^{(1)}\rangle\!\rangle +
\frac{7}{11}\, \langle\!\langle Q^{(3)}\rangle\!\rangle + 
\frac{14}{3}\, \langle\!\langle Q^{(5)}\rangle\!\rangle,\nonumber\\
\wt{s}_{10} & = & \phantom{-}\frac{3}{22}\, a_2^\perp - \frac{1}{8}\, \zeta_3
\omega_3^T - \frac{28}{55}\, \langle\!\langle Q^{(1)}\rangle\!\rangle -
\frac{7}{11}\, \langle\!\langle Q^{(3)}\rangle\!\rangle 
+ \frac{14}{3}\, \langle\!\langle Q^{(5)}\rangle\!\rangle,\nonumber\\
s_{01} & = & \phantom{-}\frac{3}{44}\, a_2^\perp + \frac{1}{8}\, \zeta_3
\omega_3^T + \frac{49}{110}\, \langle\!\langle Q^{(1)}\rangle\!\rangle -
\frac{7}{22}\, \langle\!\langle Q^{(3)}\rangle\!\rangle 
+ \frac{7}{3}\, \langle\!\langle Q^{(5)}\rangle\!\rangle,\nonumber\\
\wt s_{01} & = & -\frac{3}{44}\, a_2^\perp + \frac{1}{8}\, \zeta_3
\omega_3^T - \frac{49}{110}\, \langle\!\langle Q^{(1)}\rangle\!\rangle +
\frac{7}{22}\, \langle\!\langle Q^{(3)}\rangle\!\rangle
 + \frac{7}{3}\, \langle\!\langle Q^{(5)}\rangle\!\rangle,\nonumber\\
t_{10} & = & -\frac{9}{44}\, a_2^\perp - \frac{3}{16}\, \zeta_3
\omega_3^T - \frac{63}{220}\, \langle\!\langle Q^{(1)}\rangle\!\rangle +
\frac{119}{44}\, \langle\!\langle Q^{(3)}\rangle\!\rangle,\nonumber\\
\wt t_{10} & = & \phantom{-}\frac{9}{44}\, a_2^\perp - \frac{3}{16}\, \zeta_3
\omega_3^T + \frac{63}{220}\, \langle\!\langle Q^{(1)}\rangle\!\rangle +
\frac{35}{44}\, \langle\!\langle Q^{(3)}\rangle\!\rangle.\label{eq:CO}
\end{eqnarray}
The above relations involve the three parameters 
$\langle\!\langle Q^{(1)}\rangle\!\rangle$, 
$\langle\!\langle Q^{(3)}\rangle\!\rangle$
and $\langle\!\langle Q^{(5)}\rangle\!\rangle$, 
which can be defined as matrix elements of the
following operators:
\begin{eqnarray}
Q^{(1)}_{\alpha,\xi\eta} & = & -i\bar u \dnabla_\alpha
(\sigma_{\xi\rho} g G_{\eta\rho} - \sigma_{\eta\rho} g G_{\xi\rho} )
d +7 \bar u {\cal D}_\alpha g [G-i\gamma_5\widetilde{G}]_{\xi\eta} d -
\frac{11}{3}\,\partial_\alpha \bar u
g[G-i\gamma_5\widetilde{G}]_{\xi\eta} d,\nonumber\\
Q^{(3)}_{\alpha,\xi\eta} & = & \left\{ i\bar u  \dnabla_\xi
  \sigma_{\eta\rho} g[ G+i\gamma_5 \widetilde{G}]_{\alpha\beta}\, d -
  \frac{1}{3}\,\bar u {\cal D}_\alpha g[G+i\gamma_5\widetilde{G}]_{\xi\eta}\, d
  - \frac{1}{3}\,\bar u {\cal D}_\eta g[G+i\gamma_5\widetilde{G}]_{\xi\alpha}\,
  d \right.\nonumber\\
& & \left.{}+ \frac{1}{3}\, \partial_\alpha \bar u
  g[G+i\gamma_5\widetilde{G}]_{\xi\eta}\, d + \frac{1}{3}\, \partial_\eta
  \bar u g[G+i\gamma_5\widetilde{G}]_{\xi\alpha}\, d\right\} - \{ \xi
\leftrightarrow \eta \},\nonumber\\
Q^{(5)}_{\alpha,\xi\eta} & = & \bar u {\cal D}_\alpha g [ G + i
\gamma_5\widetilde{G}]_{\xi\eta} d - \frac{1}{2}\, \partial_\alpha
\bar u g [ G + i \gamma_5\widetilde{G} ]_{\xi\eta} d,\label{eq:defopQ}
\end{eqnarray}
where we used convenient short-hand notations (see \cite{BBK}) for
the covariant derivatives: $\dnabla_\alpha G_{\mu\nu}\equiv  
G_{\mu\nu}\derright_\alpha-\derleft_\alpha G_{\mu\nu}$ 
acting on quark fields only, and ${\cal D}_\alpha G_{\mu\nu}\equiv
[D_\alpha, G_{\mu\nu}]$ acting on gluon fields only. The reduced matrix 
elements $\langle\!\langle Q^{(i)}\rangle\!\rangle$
 of these operators are defined as
\begin{eqnarray}
\langle 0 | Q^{(i)}_{\alpha,\xi\eta} | \rho^-(P,\lambda)\rangle & = &
\left[ e^{(\lambda)}_\xi \left(P_\alpha P_\eta - \frac{1}{3}\,
    m_\rho^2 g_{\alpha\eta} \right) - e^{(\lambda)}_\eta \left
    ( P_\alpha P_\xi - \frac{1}{3}\, m_\rho^2 g_{\alpha\xi}\right)
\right] f_\rho^T m_\rho^2  
\langle\!\langle Q^{(i)}\rangle\!\rangle\nonumber\\
& & {} + (e^{(\lambda)}_\xi
g_{\alpha\eta} - e^{(\lambda)}_\eta g_{\alpha\xi} ) 
\langle\!\langle R^{(i)}\rangle\!\rangle,\label{eq:defmeQ}
\end{eqnarray}
where $ \langle\!\langle Q^{(i)}\rangle\!\rangle$ is of twist-4 and
$\langle\!\langle R^{(i)}\rangle\!\rangle$ of twist-5.
\begin{table}
\renewcommand{\arraystretch}{1.4}
\addtolength{\arraycolsep}{3pt}
$$
\begin{array}{|cccc|}\hline
f_\rho^T\,[{\rm MeV}] & a_2^\perp & \zeta_3 & \omega_3^T \\ \hline
160\pm 10 & 0.20\pm 0.10 & 0.032\pm 0.010 & 7.0 \pm 7.0 \\ \hline
\end{array}
$$
\caption[]{Parameters of twist-2 and twist-3 
           chiral-odd distribution amplitudes.
  Renormalization scale is $\mu = 1\,$GeV.}\label{tab:para2}
$$
\begin{array}{|ccccc|}\hline
 \zeta_4^T & \tilde{\zeta_4^T} & \langle\!\langle Q^{(1)}\rangle\!\rangle
 & \langle\!\langle Q^{(3)}\rangle\!\rangle & \langle\!\langle Q^{(5)}\rangle
\!\rangle\\ \hline
 0.10\pm 0.05 & -0.10\pm 0.05 & -0.15\pm 0.15 & 0 & 0\\ \hline
\end{array}
$$
\caption[]{Parameters of twist-4
           chiral-odd distribution amplitudes.
  Renormalization scale as above.}\label{tab:para3}
\renewcommand{\arraystretch}{1}
\addtolength{\arraycolsep}{-3pt}
\end{table}
The operators renormalize multiplicatively and their one-loop
anomalous dimensions are known \cite{BBK}; to obtain the scale-dependence 
of the matrix elements $\langle\!\langle Q^{(i)}\rangle\!\rangle$,
one has to subtract the anomalous dimension of $f_\rho^T$, 
Eq.~(\ref{eq:fTscaling}), so that
\begin{eqnarray}
\langle\!\langle Q^{(i)}\rangle\!\rangle (Q^2) & = &
L^{\gamma_{Q^{(i)}}/b}\, 
\langle\!\langle Q^{(i)}\rangle\!\rangle(\mu^2)\nonumber\\
\gamma_{Q^{(1)}} &=& -4C_F + \frac{11}{2}\, C_A,\quad 
\gamma_{Q^{(3)}} = \frac{10}{3}\, C_F,\quad
 \gamma_{Q^{(5)}} = -\frac{5}{3}\, C_F + 5 C_A.
\end{eqnarray}
Numerical estimates for these matrix elements were obtained in
\cite{BBK} and App.~C and are collected in Table~\ref{tab:para3}.

Finally, we have to specify the two-particle twist-4 distributions 
$h_3$ and ${\Bbb A}_T$ defined 
in Sec.~2. They are not independent, but can be expressed 
in terms of  $S$, $T_2^{(4)}$ and $T_4^{(4)}$ by using the EOM, see
Eqs.~(\ref{eq:rel1odd}),
(\ref{eq:rel2odd}). To next-to-leading accuracy, we obtain: 
\begin{eqnarray}
h_3(u) & = & 1 + \left\{ -1+\frac{3}{7}\,a_2^\perp - 10 (\zeta_4^T +
  \wt{\zeta}_4^T ) \right\} C_2^{1/2}(\xi)\nonumber\\
& & {} + \left\{
   -\frac{3}{7}\, a_2^\perp - \frac{15}{8}\, \zeta_3
  \omega_3^T\right\} C_4^{1/2}(\xi),\label{eq:h3exp}\\
{\Bbb A}_T(u) & = & 30 u^2 \bar u^2 \left\{ \frac{2}{5} \left( 1 +
  \frac{2}{7}\, a_2^\perp + \frac{10}{3}\, \zeta_4^T - \frac{20}{3}\,
  \wt{\zeta}_4^T \right) + \left( \frac{3}{35}\, a_2^\perp +
  \frac{1}{40}\, \zeta_3 \omega_3^T \right) C_2^{5/2}(\xi)
  \right\}\nonumber\\
 & & {} - \left( \frac{18}{11}\, a_2^\perp   - \frac{3}{2}\, \zeta_3
  \omega_3^T + \frac{126}{55}\,
 \langle\!\langle Q^{(1)}\rangle\!\rangle + \frac{70}{11}\, 
\langle\!\langle Q^{(3)}\rangle\!\rangle
   \right)\nonumber\\
 & & \times \left( u\bar u (2+13 u\bar u) + 2u^3 (10-15u+6u^2) \ln u +
  2\bar u^3 (10-15\bar u + 6\bar u^2) \ln \bar u\right).
\label{eq:ATexp}
\end{eqnarray}
\begin{figure}
\centerline{\epsffile{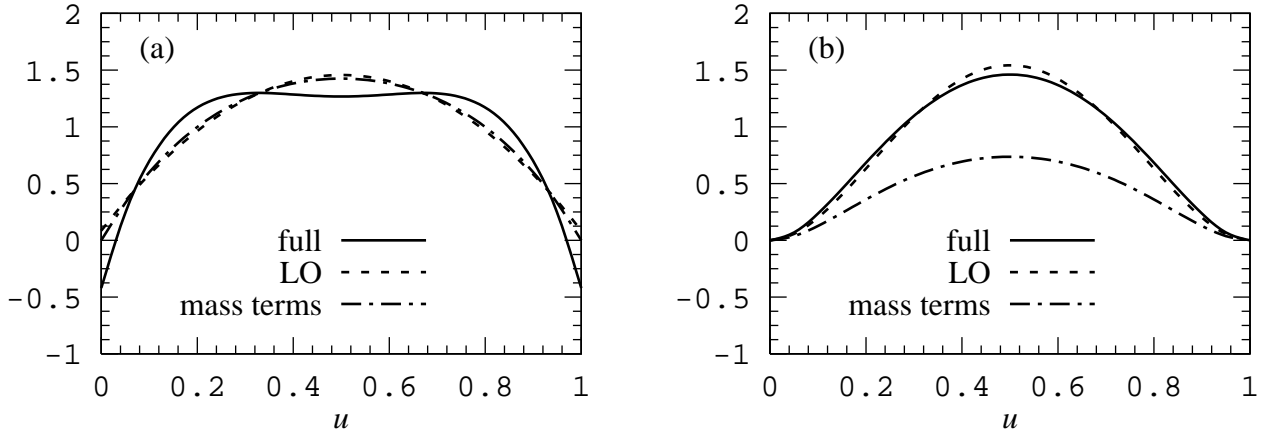}}
\caption[]{
Two-particle twist-4 chiral-odd distribution amplitudes of the $\rho$
meson: $h_3$ (a) and ${\Bbb A}_T$ (b). LO means neglecting contributions
of higher conformal spin for twist-3 and twist-4 operators and 
the mass terms correspond to retaining  meson mass corrections only.
}
\end{figure}
In Fig.~2, we plot $h_3$ and ${\Bbb A}_T$ as functions of $u$, showing
full results and the contributions from leading order conformal spin and mass
correction terms separately. Like in the chiral-even case, the mass
terms dominate $h_3(u)$ and constitute
approximately one half of ${\Bbb A}_T(u)$.

We stress that the given expressions are exact provided the three-particle 
distributions are taken in the above approximation. This means, 
in particular, that (\ref{eq:h3exp}) and (\ref{eq:ATexp}) reproduce the 
exact second moments of $h_3$ and $d^2/du^2 {\Bbb A}_T$, i.e.\ the
normalization of ${\Bbb A}_T$, but the fourth moment of $h_3$
(second of ${\Bbb A}_T$) also includes (uncalculated) contributions 
from even higher conformal spin operators. We have checked that the
second moments agree with those obtained from Taylor expanding
(\ref{eq:OPE2}). 

Note that, like $g_3$, $h_3$ corresponds to the
spin projection $s=-1/2$ for both the quark and the antiquark, 
and thus has a
conformal expansion in Gegenbauer polynomials $C^{1/2}(2u-1)$, 
cf.\ (\ref{eq:asymptotic}):
$$
h_3(u,\mu^2) = 1 + \sum_{k=2,4,\ldots}^\infty h^{(k)}_3(\mu^2)
C^{1/2}_{k}(2u-1).
$$ 
The coefficients $h^{(2)}_3$ and $h^{(4)}_3$ can be read off
(\ref{eq:h3exp}). The conformal expansion of ${\Bbb A}_T$ is 
more complicated.

\section{Summary and Conclusions}
\setcounter{equation}{0}

In the present paper we have studied the twist-4 two- and
three-particle distribution amplitudes of vector mesons in QCD and
expressed them in a model-independent way by a minimal number of
nonperturbative parameters. The work reported here is an extension of
our earlier paper on twist-3 distribution amplitudes \cite{BBKT}. The
one ingredient in the approach is the use of the QCD equations of
motion, which allow us to reveal interrelations between different
distribution amplitudes of a given twist and to obtain exact integral
representations for distribution amplitudes that are not dynamically
independent. The other ingredient is the
use of conformal expansion: analogously to partial wave
decomposition in quantum mechanics, it allows one to separate
transverse and longitudinal variables in the wave function.
 The dependence on transverse coordinates is represented as scale-dependence
of the relevant operators and is governed by
renormalization-group equations; the dependence on the longitudinal
momentum fraction is described in terms of irreducible
representations of the corresponding symmetry group, the collinear
conformal group SL(2,R). The conformal partial wave expansion is
explicitly consistent with the equations of motion since the latter are
not renormalized. The expansion thus makes maximum use of the symmetry
of the theory in order to simplify the dynamics, which is related,
in the perturbative domain, to renormalization properties of the relevant
operators. 

The analysis of twist-4 distribution amplitudes is complicated by the
fact that the twist-4 terms are of different origin: there are, first,
``intrinsic'' twist-4 corrections from matrix elements of twist-4
operators. There are, second, admixtures of matrix elements of twist-3
operators, as the counting of twist in terms of ``good'' and ``bad''
projections on light-cone coordinates does not exactly match the
definition of twist as ``dimension minus spin'' of an operator. There
are, third, meson mass corrections, which one may term  kinematical
corrections, that come, on the one hand,  from the subtraction of traces in the
leading twist operators and, on the other hand, 
from higher twist operators containing 
total derivatives of twist-2 operators. 
Meson mass corrections of the first kind are formally 
analogous to Nachtmann corrections in inclusive processes, while 
the contribution of  operators with total derivatives 
is a specific new feature of exclusive 
processes, which makes the structure of these corrections much more 
complex.

Our final results
are collected in Secs.~4 and 5. We  present a complete set of 
distribution amplitudes that is consistent with the QCD equations of motion 
and has a minimum number of nonperturbative 
parameters whose  numerical values  are estimated  
from QCD sum rules. It turns out that the
meson mass corrections are the dominant ones in all two-particle
twist-4 distributions, which
is in contrast to what is observed in deep-inelastic scattering and
welcome from the phenomenological point of view, as the higher twist
matrix elements, $\zeta_{3,4}$ etc.,
come with considerable numerical uncertainties.

The results of our study are immediately applicable --- and, in fact,
have already been applied \cite{BB98} --- to processes such as exclusive
or radiative $B$ decays and hard electroproduction of vector mesons at
HERA.

\subsection*{Acknowledgements}
P.B.\ is supported by DFG through a Heisenberg fellowship. She also
acknowledges hospitality and financial support from NORDITA during a
visit there when this work was initiated. We thank G. Stoll for
collaboration on part of the subjects covered in this work.

\appendix
\renewcommand{\theequation}{\Alph{section}.\arabic{equation}}
\setcounter{table}{0}
\renewcommand{\thetable}{\Alph{table}}

\section*{Appendices}

\section{Equations of Motion}
\label{app:a}
\setcounter{equation}{0}

\subsection{Operator Identities}

In this appendix we  collect exact operator identities, which can be 
derived using the approach of \cite{BB89} and which present a 
nonlocal equivalent to the equations of motion for Wilson local operators.
The basic idea is to study the response of nonlocal operators 
to total translations and/or the change of the interquark separation
along the light-cone.
For convenience, we work in the Fock--Schwinger gauge $x_\mu A_\mu (x)=0$, 
so that
$$
[x,-x] = 1,\qquad A_\mu(x) = \int_0^1 dv\, v x_\alpha G_{\alpha\mu}(v
x).
$$
All operator relations can be made manifestly gauge invariant by restoring the
path-ordered gauge factor between field operators at different points
in space-time.  

For the general two-particle operator, we can write
\begin{equation}
\frac{\partial}{\partial x_\mu}\, \bar u(x)\Gamma d(-x) = -\bar
u(x)\Gamma \deriv_\mu d(-x) - i \int_{-1}^1 dv\, v \bar u (x) x_\alpha
gG_{\alpha\mu}(vx) \Gamma d(-x),\label{eq:app1}
\end{equation}
where $\Gamma$ is  an arbitrary Dirac matrix and
$\deriv_\mu = \derright_\mu - \derleft_\mu = 
(\stackrel{\rightarrow}{\partial} -iA(-x))_\mu-
(\stackrel{\leftarrow}{\partial} +iA(x))_\mu$. The derivatives act
on the arguments of the quark operators. 

In a similar way we can calculate the derivative with respect to the 
total translation:
\begin{equation}\label{eq:app2}
\partial_\mu \{\bar u(x)\Gamma d(-x)\} 
= \bar u(x) (\derleft_\mu + \derright_\mu
) \Gamma d(-x) - i\int_{-1}^1 dv\, \bar u(x) x_\alpha
G_{\alpha\mu}(vx) \Gamma d(-x), 
\end{equation}
where, by definition, 
\begin{equation}
\partial_\mu \left\{ \bar u(x)\Gamma d(-x)\right\} \equiv
\left.\frac{\partial}{\partial y_\mu}\,\left\{ \bar u(x+y) [x+y,-x+y]
    \Gamma d(-x+y)\right\}\right|_{y\to 0}.
\end{equation}
Here it is important to keep the gauge factors, which give a
nonvanishing contribution:
$$
\partial_\mu [x,-x] = iA_\mu(x) [x,-x] - i [x,-x] A_\mu(-x) - i
\int_{-1}^1 dv\, [x,vx] x_\alpha G_{\alpha\mu}(vx) [vx,-x].
$$
For chiral-even operators, 
$\Gamma = \{\gamma_\mu,\gamma_\mu \gamma_5\}$, the first terms on the 
right-hand side of Eqs.~(\ref{eq:app1}), (\ref{eq:app2})
vanish by virtue of the massless Dirac equation, so that
\begin{eqnarray}
\frac{\partial}{\partial x_\mu}\, \bar u(x)\gamma_\mu(\gamma_5) d(-x)
& = &{} - i \int_{-1}^1 dv\, v \bar u (x) x_\alpha
gG_{\alpha\mu}(vx) \gamma_\mu(\gamma_5) d(-x),\label{eq:oprel1}\\
\partial_\mu \{\bar u(x)\gamma_\mu(\gamma_5) d(-x)\} 
& = & {} - i\int_{-1}^1 dv\, \bar u(x) x_\alpha
G_{\alpha\mu}(vx) \gamma_\mu(\gamma_5) d(-x).\label{eq:oprel2}
\end{eqnarray}
For chiral-odd operators, 
 $\Gamma = \{{\bf 1}(\gamma_5),\sigma_{\mu\nu}(\gamma_5)\}$, 
on the other hand, we can use the identities
\begin{equation}
\sigma_{\mu\nu}\deriv_\mu = i (\derright_\nu + \derleft_\nu),\qquad 
(\derright_\mu + \derleft_\mu)\sigma_{\mu\nu} = i \deriv_\nu,
\end{equation}
and get, combining Eqs.~(\ref{eq:app1}) and (\ref{eq:app2}):
\begin{eqnarray}
\partial_\mu \bar u(x) \sigma_{\mu\nu}(\gamma_5) d(-x) & = &
-i\,\frac{\partial\phantom{x_\nu}}{\partial x_\nu} \,\bar u(x)(\gamma_5)
d(-x) + \int_{-1}^1 dv\, v \bar u(x)
x_\rho gG_{\rho\nu}(vx)(\gamma_5)d(-x)\nonumber\\
& & {} -i\int_{-1}^1 dv\, \bar u(x) x_\rho gG_{\rho\mu}(vx) 
\sigma_{\mu\nu} (\gamma_5)d(-x),\label{eq:rel1}\\
\frac{\partial\phantom{x_\nu}}{\partial x_\mu} \,\bar u(x)
\sigma_{\mu\nu}(\gamma_5) d(-x) & = & -i\partial_\nu \bar u(x)
(\gamma_5)d(-x) + 
\int_{-1}^1 dv\, \bar u(x) x_\rho gG_{\rho\nu}(vx)(\gamma_5)d(-x)\nonumber\\
& & {} -
i\int_{-1}^1 dv\, v \bar u(x) x_\rho gG_{\rho\mu}(vx) 
\sigma_{\mu\nu}(\gamma_5) d(-x).
\label{eq:rel2}
\end{eqnarray}
This method is general and can also be used for calculating the second 
derivative. In particular, the following formula is useful:
\begin{eqnarray}
\lefteqn{
\frac{\partial^2}{\partial x_\alpha \partial x^\alpha} \bar u(x)\Gamma\, d(-x)
 = -\partial^2 \bar u(x)\Gamma\, d(-x) +\bar u(x) 
[\Gamma\sigma G+\sigma G\,\Gamma]d(-x)}
\nonumber\\
&&{}-2ix^\nu \frac{\partial}{\partial x_\mu}\!\int_{-1}^1\!\!dv\,v\,\bar u(x)
\Gamma G_{\nu\mu}(vx)d(-x)
-2ix^\nu \partial_\mu\! \int_{-1}^1\!\!dv\,\bar u(x)
\Gamma G_{\nu\mu}(vx)d(-x)
\nonumber\\
&&{}+2\int_{-1}^1\!dv\int_{-1}^v\!dt\,(1+vt)\bar u(x)\Gamma x^\mu x^\nu
G_{\mu\rho}(vx)G^{\rho}_{\phantom{\nu}\nu}(tx)d(-x)
\nonumber\\
&&{}+ix^\nu \int_{-1}^1\!\!dv\,(1+v^2)\,\bar u(x)\Gamma 
[D_\mu,G^{\mu}_{\phantom{\nu}\nu}](vx) d(-x),
\label{twoderiv}
\end{eqnarray}
where 
$[D_\mu,G^{\mu}_{\phantom{\nu}\nu}] =-t^A(\bar\psi\gamma_\nu t^A\psi)$ ,
assuming  summation over light flavours $\psi$.
 
\subsection{Relations Between Distribution Amplitudes}

We are now in a position to derive relations between two- and
three-particle amplitudes. 
Sandwiching (\ref{eq:oprel1}) between the
  vacuum and the $\rho$ meson state, we find
\begin{equation}
g_3(u) = \phi_\parallel(u) -2 \,\frac{d\phantom{u}}{du} \int_0^u
d\alpha_d \int_0^{\bar u} d\alpha_u \,\frac{1}{\alpha_g} \,\left\{ 2
  \Phi(\underline{\alpha}) + \Psi(\underline{\alpha}) \right\}.
\label{eq:rel1even}\end{equation}
To arrive at Eq.~(\ref{eq:rel1even}), the following formula 
proves useful:
\begin{eqnarray}
\lefteqn{\int_{-1}^1 dv\, v^k \int {\cal D}\alpha\, {\cal
  F}(\underline{\alpha}) \exp[ -i px \{ \alpha_u-\alpha_d+v
  \alpha_g\}] \ =}\hspace*{1.5cm}\nonumber\\
& = & \int_0^1 du\, \exp[ i\xi px] \int_0^u d\alpha_d \int_0^{\bar u}
  d\alpha_u \,\frac{2}{\alpha_g}\,\left[ \frac{1}{\alpha_g}\,
  (\alpha_d-\alpha_u-\xi)\right]^k {\cal F}(\underline{\alpha}),
\end{eqnarray}
which is valid for an arbitrary function ${\cal F}(\underline{\alpha})$. 

On the other hand, taking the matrix element of (\ref{eq:oprel2}), 
and eliminating $g_3$ by virtue of (\ref{eq:rel1even}), we obtain
\begin{eqnarray}
\frac{1}{8}\,\frac{d^2}{du^2}\,{\Bbb A}(u) & = & 4
(g_\perp^v(u)-\phi_\parallel(u)) + 4 \,\frac{d}{du} \int_0^u
d\alpha_d \int_0^{\bar u} d\alpha_u \,\frac{1}{\alpha_g} \left\{ 2
  \Phi(\underline{\alpha}) + \Psi(\underline{\alpha}) \right\}
\nonumber\\
& & {}+ \frac{d^2}{du^2}\int_0^u
d\alpha_d \int_0^{\bar u} d\alpha_u \,\frac{1}{\alpha_g^2}\, (\alpha_u
u -\alpha_d \bar u) \left\{ 2
  \Phi(\underline{\alpha}) + \Psi(\underline{\alpha}) \right\}.
\label{eq:rel2even}
\end{eqnarray}
${\Bbb A}$ can then be obtained as
$$
{\Bbb A}(u) = \int_0^u dv \int_0^v dw\, \frac{d^2}{dw^2}\,{\Bbb A}(w).
$$

Two more relations are derived in a similar manner 
between chiral-odd distribution amplitudes using the 
operator identities in (\ref{eq:rel1}) and 
(\ref{eq:rel2}). {}From  (\ref{eq:rel1}):
\begin{eqnarray}
\lefteqn{h^{(s)}_\parallel(u)-
\frac{1}{2}\,(h_3(u)+\phi_\perp(u))\ =}\hspace*{1.4cm}\nonumber\\
& = &\frac{d\phantom{u}}{du}
\int_0^u\!\! d\alpha_d \int_0^{\bar u}\!\! d\alpha_u \left
  [ \frac{\alpha_d-\alpha_u-\xi}{\alpha_g^2} \,{
    S}(\underline{\alpha})
-\frac{1}{\alpha_g}\left
    ( T_2^{(4)}(\underline{\alpha}) -
    T_3^{(4)}(\underline{\alpha})\right)
\right],\hspace*{10pt}\label{eq:rel1odd} 
\end{eqnarray}
and from (\ref{eq:rel2}): 
\begin{eqnarray}
\lefteqn{4 \int_0^udv \int_0^v dw \, [h_3(w)-\phi_\perp(w)] 
-\frac{1}{2}\,{\Bbb A}_T(u) -\int_0^u
  dv\,(2v-1) [\phi_\perp(v)
+h_3(v)]\ =}\hspace*{2cm}\nonumber\\
& = & \int_0^u d\alpha_d \int_0^{\bar u} d\alpha_u
\,\frac{2}{\alpha_g} \,\left\{\frac{\alpha_d-\alpha_u-\xi}{\alpha_g}\left(
T_2^{(4)}(\underline{\alpha}) -
T_3^{(4)}(\underline{\alpha})\right) - {
  S}(\underline{\alpha})\right\}.\hspace*{2cm} \label{eq:rel2odd}
\end{eqnarray}

\section{Short-Distance Expansion of Distribution 
Am\-pli\-tu\-des: Relation to Local Operators}\label{app:B}
\setcounter{equation}{0}

In this appendix we calculate the next-to-leading 
corrections to the conformal expansion of 
twist-4 three-particle distribution amplitudes as given
in Eq.~(\ref{eq:NLO}) and (\ref{eq:CO}).

\subsection{Chiral-Even}

Our general strategy will be to consider matrix elements of the
relevant operators with all  Lorentz indices open. Taking 
different light-cone projections we will relate coefficients 
in the conformal expansion of distribution amplitudes to the 
invariant Lorentz structures and then 
identify the relevant contractions of indices that  
relate $\phi_{10}$, $\phi_{01}$, $\psi_{10}$ and $\wt{\psi}_{10}$
to matrix elements of independent twist-4 operators.
  
To the next-to-leading conformal spin accuracy, we need local 
operators of dimension 6 with one quark-antiquark pair, 
one gluon field and one additional covariant derivative.
Taking into account G-parity, only two operators can contribute, apart
from operators with total derivatives:
\begin{eqnarray}
 O^{(1)}_{\alpha\beta\mu\nu} & = & \bar u (i\derleft_\beta
 g\wt{G}_{\mu\nu} + g\wt{G}_{\mu\nu}
 i\derright_\beta)\gamma_\alpha\gamma_5 d,\nonumber\\
 O^{(2)}_{\alpha\beta\mu\nu} & = & \bar u (-\derleft_\beta
 g{G}_{\mu\nu} + g{G}_{\mu\nu}
 \derright_\beta)\gamma_\alpha d.
\end{eqnarray}
For each of them we write down a general Lorentz decomposition:
\begin{eqnarray}
\langle 0 | O^{(i)}_{\alpha\beta\mu\nu} | \rho^-(P,\lambda)
\rangle & = & \left\{ e^{(\lambda)}_\mu \left[ P_\alpha P_\beta P_\nu - 
\frac{5}{24}\,
  m_\rho^2 (P_\alpha g_{\beta\nu} + P_\beta g_{\alpha\nu}) -
  \frac{1}{6}\, m_\rho^2 P_\nu g_{\alpha\beta}\right]\right.
\nonumber\\
&&  {} - 
e^{(\lambda)}_\nu \left[ P_\alpha P_\beta P_\mu - \frac{5}{24}\,
  m_\rho^2 (P_\alpha g_{\beta\mu} + P_\beta g_{\alpha\mu}) -
  \frac{1}{6}\, m_\rho^2 P_\mu g_{\alpha\beta}\right]\nonumber\\
&& \left. {} -
\frac{1}{24}\, m_\rho^2 \left[ e^{(\lambda)}_\alpha (g_{\beta\nu} P_\mu -
  g_{\beta\mu} P_\nu) + e^{(\lambda)}_\beta (g_{\alpha\nu} P_\mu - 
g_{\alpha\mu}
  P_\nu)\right] \right\} A^{(i)} f_\rho m_\rho\nonumber\\
& & {}+ P_\alpha (e^{(\lambda)}_\mu g_{\beta\nu} -e^{(\lambda)}_\nu 
g_{\beta\mu}) B^{(i)} +
  P_\beta (e^{(\lambda)}_\mu g_{\alpha\nu} -e^{(\lambda)}_\nu
  g_{\alpha\mu}) C^{(i)}\nonumber\\
& & {}+ e^{(\lambda)}_\alpha (P_\mu g_{\beta\nu} -P_\nu g_{\beta\mu}) D^{(i)} +
  e^{(\lambda)}_\beta (P_\mu g_{\alpha\nu} - P_\nu g_{\alpha\mu})
  E^{(i)}\nonumber\\
& & -g_{\alpha\beta} (P_\mu e^{(\lambda)}_\nu - P_\nu
e^{(\lambda)}_\mu) F^{(i)}.\label{eq:horror}
\end{eqnarray}
Here $A^{(i)}$ is of twist-3 and can easily be related to an
  integral over the 
  twist-3 distribution amplitudes ${\cal A}$ and ${\cal V}$,
  respectively.  Using (\ref{eq:4.7}) and (\ref{eq:4.8}), we find
\begin{eqnarray}
-(pz)^3 e^{(\lambda)}_\perp m_\rho f_\rho 
A^{(1)} & = & \langle 0 | O^{(1)}_{\cdot\cdot\cdot\perp} |
\rho\rangle\ =\ -(pz)^3 e^{(\lambda)}_\perp m_\rho f_\rho \zeta_3
\left( \frac{3}{7} +\frac{3}{28}\, \omega_3^A\right),\label{eq:A1}\\
(pz)^3 e^{(\lambda)}_\perp m_\rho f_\rho\,
A^{(2)} & = & \langle 0 | O^{(2)}_{\cdot\cdot\cdot\perp} |
\rho\rangle\ =\ (pz)^3 e^{(\lambda)}_\perp m_\rho f_\rho\,
\frac{3}{28}\,\zeta_3\omega_3^V.\label{eq:A2}
\end{eqnarray}

To project onto the intrinsic twist-4 contributions, we must
replace one ``dot'' projection by a ``perp'' projection in
(\ref{eq:A1}) and (\ref{eq:A2}), which yields
\begin{eqnarray}
\langle 0 | O^{(1)}_{\perp\cdot\cdot\perp} | \rho\rangle & = & f_\rho
m_\rho^3 (ez) (pz) g^\perp_{\perp\perp} \left\{ \frac{1}{2} \left
    ( -\frac{1}{3}\,\zeta_3 + \frac{1}{3}\,\zeta_4 \right) -
  \frac{1}{14}\, (\phi_{01}+\phi_{10}) \right\},\nonumber\\
\langle 0 | O^{(2)}_{\perp\cdot\cdot\perp} | \rho\rangle & = & f_\rho
m_\rho^3 (ez) (pz) g^\perp_{\perp\perp} \left\{ \frac{1}{6} \left
    ( -\frac{1}{3}\,\zeta_3 + \frac{1}{3}\,\zeta_4 \right) +
  \frac{1}{14}\, (\phi_{01}-\phi_{10}) \right\},\label{eq:T4contr}
\end{eqnarray}
whereas direct contraction of (\ref{eq:horror}) gives
\begin{equation}\label{eq:B.6}
\langle 0 | O^{(i)}_{\perp\cdot\cdot\perp} | \rho\rangle =
-\frac{1}{4}\, m_\rho^3 f_\rho (ez)(pz) g^\perp_{\perp\perp} A^{(i)} +
(ez)(pz) g^\perp_{\perp\perp} \left( C^{(i)} + E^{(i)} \right). 
\end{equation}
Once $C^{(i)} + E^{(i)}$ are known in terms of $\zeta_3$, $\zeta_4$
and $\omega_{3,4}^{V,A}$, these two equations serve to determine 
$\phi_{01}$ and $\phi_{10}$.

In order to determine  $C^{(i)} + E^{(i)}$, 
we first have to introduce some more matrix elements:
\begin{eqnarray}
\langle 0 | O^{(i)}_{\xi\beta\xi\nu} | \rho\rangle & = & (e_\beta
P_\nu + e_\nu P_\beta) X_+^{(i)} + (e_\beta
P_\nu - e_\nu P_\beta) X_-^{(i)},\nonumber\\
\langle 0 | O^{(i)}_{\alpha\xi\xi\nu} | \rho\rangle & = & (e_\alpha
P_\nu + e_\nu P_\alpha) Y_+^{(i)} + (e_\alpha
P_\nu - e_\nu P_\alpha) Y_-^{(i)}.
\end{eqnarray}
By construction, $X_+$ and $Y_+$ are of twist-4 and $X_-$ and $Y_-$ of
twist-5; thus, to our accuracy:
$$
X^{(i)}_- = Y^{(i)}_- = 0.
$$
Also note that, by definition, Eq.~(\ref{eq:w4A}):
\begin{equation}
X_+^{(1)} = \zeta_4 \left(\omega_4^A - \frac{5}{18}\right).\label{eq:X}
\end{equation}
Now, by contracting (\ref{eq:horror}) with $g_{\alpha\beta}$, etc., we
find a set of linear equations relating $A$, $B$, \dots to $X_+$ and
$Y_+$, which can be solved to give
\begin{eqnarray}
B & = & D = \phantom{-}\frac{1}{4}\left( X_+-3 Y_+\right) f_\rho
m_\rho^3,\nonumber\\
C & = & E  =  -\frac{1}{4}\left( X_+-3 Y_+\right) f_\rho
m_\rho^3,\nonumber\\
F & = & 0.\label{eq:B.9}
\end{eqnarray}
We recall that contributions from twist-5 operators
are neglected in these solutions.

{}Of the remaining  three unknowns  $X_+^{(2)}$, $Y_+^{(1)}$  and
$Y_+^{(2)}$,  $Y_+^{(1)}$ can be obtained rather easily by observing
that $D_\mu \wt{G}_{\mu\nu}=0$, so that
\begin{eqnarray*}
\langle 0 | O^{(1)}_{\alpha\xi\xi\nu} + O^{(1)}_{\nu\xi\xi\alpha} |
\rho\rangle & = & P_\xi \langle 0 | \bar u
(\wt{G}_{\xi\nu}\gamma_\alpha \gamma_5 + \wt{G}_{\xi\alpha}\gamma_\nu
\gamma_5 ) d | \rho\rangle\\
 & = & - (e_\nu P_\alpha + e_\alpha P_\nu) f_\rho m_\rho^3 \left
   ( \frac{2}{3}\, \zeta_3 + \frac{1}{3}\,\zeta_4\right),
\end{eqnarray*}
which means
\begin{equation}\label{eq:B.10}
Y_+^{(1)} = -\frac{1}{3}\,\zeta_3 - \frac{1}{6}\,\zeta_4.
\end{equation}

In order to determine the remaining parameters $X_+^{(2)}$ and
$Y_+^{(2)}$, we make use of the operator identities
\begin{eqnarray}
O^{(2)}_{\xi\beta\xi\alpha} - O^{(2)}_{\beta\xi\xi\alpha} & = & 
O^{(1)}_{\xi\alpha\xi\beta} - O^{(1)}_{\alpha\xi\xi\beta} +
g_{\alpha\beta}\, O^{(1)}_{\sigma\xi\xi\sigma}
\end{eqnarray}
and
\begin{eqnarray}
\frac{4}{5} \partial_\mu E_{\mu\alpha\beta} & = & -12 i \bar u \gamma_\rho
\left\{G_{\rho\beta} \derright_\alpha - \derleft_\alpha G_{\rho\beta} +
  (\alpha\leftrightarrow \beta) \right\} d 
-4\partial_\rho \bar u (\gamma_\beta\wt{G}_{\alpha\rho} +
\gamma_{\alpha} \wt{G}_{\beta\rho} ) \gamma_5 d
\nonumber\\
& & {} - \frac{8}{3}\,
\partial_\beta \bar u \gamma_\sigma \wt{G}_{\sigma\alpha} \gamma_5 d
- \frac{8}{3}\,\partial_\alpha \bar u \gamma_\sigma
\wt{G}_{\sigma\beta} 
\gamma_5 d
+\frac{28}{3}\, g_{\alpha\beta} \partial_\rho \bar u \gamma_\sigma
\wt{G}_{\sigma\rho} d,\label{eq:twist2}
\end{eqnarray}
where 
$$E_{\mu\alpha\beta} =\left[
\frac{15}{2}\bar u \gamma_\mu \deriv_{\alpha} \deriv_{\beta}  d
-\frac{3}{2} \partial_{\alpha} \partial_{\beta}\bar u \gamma_\mu  d
  - {\rm traces}\right]_{\rm symmetrized}
$$
is a leading twist-2 conformal operator. Taking matrix elements, we find
\begin{eqnarray}
X_+^{(2)} - Y_+^{(2)} & = & X_+^{(1)} - Y_+^{(1)},
\end{eqnarray}
and
\begin{eqnarray}
\frac{4}{7}\, m_\rho^2 a_2^\parallel & = & -24
f_\rho m_\rho^3 X_+^{(2)} 
+ 4 f_\rho m_\rho^3
\left(\frac{2}{3}\, \zeta_3 + \frac{1}{3}\, \zeta_4\right) -
  \frac{8}{3}\, f_\rho m_\rho^3 \zeta_4.\label{eq:B.12}
\end{eqnarray}
With  $X^{(i)}$ and $Y^{(i)}$ from Eqs.~(\ref{eq:X}),
(\ref{eq:B.10}) and (\ref{eq:B.12}), we get $C+E$ from (\ref{eq:B.9}),
and thus $\phi_{10}$ and $\phi_{01}$ from (\ref{eq:T4contr}) and 
(\ref{eq:B.6}), see the first two lines of Eq.~(\ref{eq:NLO}).

Note that it is precisely operator relations of  type
(\ref{eq:twist2}), where the divergence of a leading twist conformal 
operator is expressed as a certain combination of quark--quark--gluon
operators, that make the analysis of meson mass corrections to 
twist-4 distribution amplitudes so complicated.  
This divergence vanishes in a free theory, as expected. 

The determination of the remaining parameters $\psi_{10}$ and
$\wt{\psi}_{10}$ is now fairly easy: introducing a different ``bad''
component in (\ref{eq:A1}) and (\ref{eq:A2}), we find
\begin{equation}
\langle 0 | O^{(i)}_{\cdot\cdot\cdot *} | \rho\rangle = f_\rho
m_\rho^3 (ez)(pz) \, \frac{1}{2} \left\{ A^{(i)} - \left( X_+^{(i)} +
    Y_+^{(i)} \right) \right\}.\label{eq:x}
\end{equation}
On the other hand, taking proper integrals over distribution
amplitudes:
\begin{eqnarray}
\langle 0 | O^{(1)}_{\cdot\cdot\cdot *} | \rho\rangle & = & (ez)(pz)
f_\rho m_\rho^3 \left( \frac{2}{3}\, \wt{\psi}_{00} - \frac{2}{21}\,
  \wt{\psi}_{10} \right),\nonumber\\
\langle 0 | O^{(2)}_{\cdot\cdot\cdot *} | \rho\rangle & = &
-\frac{2}{21}\, f_\rho m_\rho^3 (ez)(pz) \psi_{10}.\label{eq:y}
\end{eqnarray}
By equating (\ref{eq:x}) and (\ref{eq:y}), we obtain the last two
lines of Eq.~(\ref{eq:NLO}).

\subsection{Chiral-Odd}

The calculation of next-to-leading order spin corrections to
chiral-odd distribution amplitudes essentially parallels the calculation
of similar corrections to the photon distribution amplitude
in Ref.~\cite{BBK}. To follow this analogy, it is convenient to 
express the corrections in terms of 
\begin{equation}
\eta_1 = -\frac{3}{4}\, (s_{10} + 2 s_{01}),\quad \eta_2 =
-\frac{1}{4} \,(s_{10}-2s_{01})\label{eq:etas}
\end{equation}
and the corresponding ``dual'' quantities $\wt\eta_1$, $\wt\eta_2$, 
instead of the coefficients $s_{01},s_{10},\wt s_{10},\wt s_{01}$
in the expansion over Appell polynomials.
 
Expanding (\ref{eq:T3}) and (\ref{eq:2.21}) in powers of $(pz)$ to first
order, we obtain
\begin{eqnarray}
\lefteqn{\langle 0 | \bar u \dnabla_\cdot \sigma_{\alpha\beta} gG_{\mu\nu} d |
\rho^-\rangle\ =\ f_\rho^T m_\rho^2 \,\frac{ez}{2pz}\,\left[
   p_\alpha p_\mu g^\perp_{\beta\nu} - \dots\right] ipz
 \left(-\frac{3}{28}\, \zeta_3 \omega_3^T\right)}\nonumber\\
& & {} + f_\rho^T m_\rho^2 ipz \left\{ \left[ p_\alpha e^\perp_\mu
    g^\perp_{\beta\nu} - \dots\right] \frac{2}{21}\, t_{10} + \left[
    e^\perp_\alpha p_\mu g^\perp_{\beta\nu} - \dots\right] \left(
    -\frac{1}{6}\, \wt\zeta_4^T - \frac{1}{42}\,\wt\eta_1 -
    \frac{3}{14}\, \wt\eta_2\right) \right.\nonumber\\
& & \left. {} -\left[ p_\alpha p_\mu e^\perp_\beta z_\nu - \dots\right]
  \frac{1}{pz}\, \frac{2}{21} \wt t_{10} + \left[ p_\alpha p_\mu
    z_\beta e^\perp_\nu - \dots \right] \frac{1}{pz} \left(
    \frac{1}{6}\, \zeta_4^T + \frac{1}{42}\, \eta_1 + \frac{3}{14}\,
    \eta_2\right)\! \right\} + O(\mbox{\rm
  twist-5}),\nonumber\\[-15pt]\label{eq:star}\\[-40pt]\nonumber
\end{eqnarray}
\begin{eqnarray}
\langle 0 | \bar u g{\cal D}_\cdot G_{\mu\nu} d | \rho^-\rangle & = &
f_\rho^T m_\rho^2 pz (e_\mu^\perp p_\nu - e_\nu^\perp p_\mu) \left\{
  \frac{1}{2}\, \zeta_4 - \frac{1}{14}\left(
    \eta_1+\eta_2\right)\right\},\nonumber\\
\langle 0 | \bar u ig{\cal D}_\cdot \wt G_{\mu\nu} d | \rho^-\rangle & = &
f_\rho^T m_\rho^2 pz (e_\mu^\perp p_\nu - e_\nu^\perp p_\mu) \left\{
  \frac{1}{2}\, \wt\zeta_4 - \frac{1}{14}\left(
    \wt\eta_1+\wt\eta_2\right)\right\}.\label{eq:star2}
\end{eqnarray}
Comparing these expressions  with the  reduced matrix 
elements  of the conformal operators, Eq.~(\ref{eq:defmeQ}), 
we immediately find
\begin{equation}
\langle\!\langle Q^{(5)}\rangle\!\rangle
     = - \frac{1}{14}\, (\eta_1+\eta_2) - \frac{1}{14}\, (
\wt\eta_1+\wt\eta_2),\label{eq:B.18}
\end{equation}
while contracting (\ref{eq:star}) over $g_{\beta\nu}$, we find
\begin{equation}
\langle\!\langle Q^{(1)} \rangle\!\rangle
 = -\frac{2}{21}\, (t_{10} - \wt t_{10} ) - \frac{10}{21}
(\eta_1-\wt \eta_1) - \frac{2}{7}\, (\eta_2-\wt\eta_2).\label{eq:B.19}
\end{equation}
Next, we introduce one more operator 
\begin{equation}
Q^{(2)}_{\alpha,\xi\eta} = i\bar u \dnabla_\alpha (\sigma_{\xi\rho} g
  G_{\eta\rho} - \sigma_{\eta\rho} g G_{\xi\rho} ) d + \frac{1}{3}\,
  \bar u {\cal D}_\alpha (gG_{\xi\eta} - i g\wt{G}_{\xi\eta} \gamma_5 ) d.
\end{equation}
Using (\ref{eq:star}) and (\ref{eq:star2}), we easily find
\begin{equation}
\langle\!\langle Q^{(2)}\rangle\!\rangle 
    = \frac{2}{21}\, (t_{10} - \wt t_{10}) - \frac{1}{21}\, (\eta_1
- \wt \eta_1) - \frac{5}{21}\, (\eta_2 - \wt \eta_2).\label{eq:B.23}
\end{equation}
$Q^{(2)}$ is actually not independent, but related to $Q^{(3)}$
via an important operator identity derived in
\cite{BBK}:\footnote{Note that we obtain a different sign in front of
  the total derivative operator on the right-hand side as compared to
 Eq.~(4.24) in \cite{BBK}.  We thank G. Stoll for checking this
  equation.}
\begin{equation}
Q^{(2)}_{\alpha,\xi\eta} = \frac{1}{3}\, Q^{(3)}_{\alpha,\xi\eta} +
\frac{1}{30}\, \partial_\rho \left( O^2_{\xi,\eta\alpha\rho} -
  O^2_{\eta,\xi\alpha\rho} \right),\label{eq:horror2}
\end{equation}
where $O^2$ is the leading twist-2 conformal operator
\begin{equation}
O^2_{\alpha\cdot\cdot\cdot} = \frac{15}{2}\, \bar u
\sigma_{\alpha\cdot} \dnabla_\cdot \dnabla_\cdot d - \frac{3}{2}\,
\partial^2_\cdot \bar u \sigma_{\alpha\cdot} d.
\end{equation}
The matrix element of the conformal operator
  on the right-hand side of Eq.~(\ref{eq:horror2}) is equal to
\begin{eqnarray}
\langle 0 | O^2_{\xi,\eta\alpha\rho} - O^2_{\eta,\xi\alpha\rho} |
\rho^-\rangle & = & i f_\rho^T \left( -\frac{24}{7}\, a_2^\perp\right)
\left\{ P_\alpha P_\rho (e_\xi P_\eta - e_\eta P_\xi) - \frac{1}{6}\,
  m_\rho^2 g_{\alpha\rho} (e_\xi P_\eta - e_\eta
  P_\xi)\right.\nonumber\\
&&{} + \frac{1}{24}\, m_\rho^2 \left[ e_\alpha (g_{\xi\rho} P_\eta -
      g_{\eta\rho} P_\xi) + e_\rho (g_{\xi\alpha} P_\eta -
      g_{\eta\alpha} P_\xi)\right]\nonumber\\
&&\left.{} + \frac{5}{24}\, m_\rho^2 \left
      [ P_\alpha (g_{\xi\rho} e_\eta - g_{\eta\rho} e_\xi) + P_\rho
      ( g_{\xi\alpha} e_\eta - g_{\eta\alpha} e_\xi ) \right]
  \right\},
\end{eqnarray}
so that 
\begin{equation}
\langle\!\langle Q^{(2)}\rangle\!\rangle 
= \frac{1}{3}\, \langle\!\langle Q^{(3)} \rangle\!\rangle 
- \frac{1}{14}\, a_2^\perp.\label{eq:B.25}
\end{equation}

At this point we have established three  relations for the six independent
parameters.  Three more relations follow
from the analysis of the most general matrix element
\begin{eqnarray}
\langle 0 | O_{\alpha,\mu\nu,\xi\eta} | \rho^-\rangle & = & \langle 0
| \bar u \dnabla_\alpha \sigma_{\mu\nu} ig G_{\xi\eta} d |
\rho^-\rangle\nonumber\\
&=&P_\alpha \left[ e_\mu (P_\xi g_{\nu\eta} - P_\eta g_{\nu\xi}) - e_\nu
  (P_\xi g_{\mu\eta} - P_\eta g_{\mu\xi}) \right] A \nonumber\\
& &{} +P_\alpha \left[ P_\mu (e_\xi g_{\nu\eta} - e_\eta g_{\nu\xi}) - P_\nu
  (e_\xi g_{\mu\eta} - e_\eta g_{\mu\xi}) \right] B \nonumber\\
&&{} + e_\alpha \left[ P_\mu (P_\xi g_{\nu\eta} - P_\eta g_{\nu\xi}) - P_\nu
  (P_\xi g_{\mu\eta} - P_\eta g_{\mu\xi}) \right] C \nonumber\\
&&{} + (g_{\alpha\mu} P_\nu - g_{\alpha\nu} P_\mu) ( e_\xi P_\eta -
e_\eta P_\xi) D + (g_{\alpha\xi} P_\eta - g_{\alpha\eta} P_\xi) 
( e_\mu P_\nu - e_\nu P_\mu) E\nonumber\\
&&{}+ e_\alpha ( g_{\mu\xi} g_{\nu\eta} - g_{\mu\eta} g_{\nu\xi} ) F + 
\left[ g_{\alpha\mu} (g_{\xi\nu} e_\eta - g_{\eta\nu} e_\xi) -
  g_{\alpha\nu} (g_{\xi\mu} e_\eta - g_{\eta\mu} e_\xi) \right] G\nonumber\\
 &&{}+ \left[ g_{\alpha\xi} (g_{\eta\mu} e_\nu - g_{\eta\nu} e_\mu) -
  g_{\alpha\eta} (g_{\xi\mu} e_\nu - g_{\xi\nu} e_\mu) \right] H.
\label{eq:horror3}
\end{eqnarray}
By projecting onto different light-cone variables, we find a set of
linear equations for the coefficients $A$,\dots, $E$ ($F$, $G$, $H$
are of twist-5 and thus not relevant for the following discussion):
\begin{eqnarray}
A+B+C & = & f_\rho^T m_\rho^2 \,\frac{3}{56}\, \zeta_3
\omega_3^T,\nonumber\\
B & = & -\frac{2}{21}\, f_\rho^T m_\rho^2 t_{10},\nonumber\\
A & = & f_\rho^T m_\rho^2 \left( \frac{1}{6}\, \wt\zeta_4^T +
  \frac{1}{42}\, \wt\eta_1 + \frac{3}{14}\,
\wt\eta_2\right),\nonumber\\
A-E & = & -\frac{2}{21}\, f_\rho^T m_\rho^2 \wt t_{10},\nonumber\\
B-D & = & f_\rho^T m_\rho^2  \left( \frac{1}{6}\,\zeta_4^T  +
  \frac{1}{42}\, \eta_1 + \frac{3}{14}\, \eta_2\right).\label{eq:LE}
\end{eqnarray}
Contracting (\ref{eq:horror3}) with $g_{\alpha\mu}$, we find
\begin{eqnarray}
\lefteqn{\langle 0 | \bar u {\cal D}_\nu G_{\xi\eta} d - \partial_\nu \bar u
G_{\xi\eta} d | \rho^-\rangle\ =\  \langle 0 |
O_{\rho,\rho\nu,\xi\eta} | \rho^-\rangle}\hspace*{1cm} \nonumber\\
& = & P_\nu [ P_\xi e_\eta - P_\eta e_\xi] (B-C-3D-E) + [e_\xi
  g_{\nu\xi} - e_\eta g_{\nu\xi}] (m_\rho^2 B + F-3 G-H).\hspace*{0.5cm}
\end{eqnarray}
Using (\ref{eq:star2}), this can be translated into
\begin{equation}
\zeta_4^T - \left( \frac{1}{2}\, \zeta_4^T - \frac{1}{14}\, (\eta_1 +
  \eta_2) \right) = \frac{2}{21}\, (t_{10} - \wt t_{10}) -
  \frac{3}{56}\,\zeta_3 \omega_3^T + 3 
\left(\frac{1}{6}\,\zeta_4^T + \frac{1}{42}\,\eta_1 + \frac{3}{14}\,
  \eta_2 \right),
\end{equation}
so that finally
\begin{equation}
\frac{2}{3}\, (t_{10} - \wt t_{10}) = \frac{3}{8}\, \zeta_3 \omega_3^T
- 4\eta_2.\label{eq:xyz}
\end{equation}
The same analysis can be performed for the matrix element of the dual
operator
\begin{equation}
\wt O_{\alpha,\mu\nu,\xi\eta} = \bar u i \dnabla_\alpha \wt
\sigma_{\mu\nu} \wt G_{\xi\eta} d,
\end{equation}
which results in the ``dual'' version of the relation (\ref{eq:xyz}):
\begin{equation}
\frac{2}{3}\, (\wt t_{10} -  t_{10}) = \frac{3}{8}\, \zeta_3 \omega_3^T
- 4\wt\eta_2.\label{eq:B.32}
\end{equation}

To obtain the last relation from which
$\eta_1,\eta_2,\wt\eta_1,\wt\eta_2,t_{10},\wt t_{10}$ can be
extracted, we use the identity:
\begin{eqnarray}
Q^{(3)}_{\alpha,\xi\eta} & = & -\frac{2}{3}\left
  ( Q^{(5)}_{\alpha,\xi\eta} - \frac{1}{2}\,\partial_\alpha
  O^+_{\xi\eta} \right) -\frac{1}{3}\left(
   Q^{(5)}_{\eta,\xi\alpha} - \frac{1}{2}\,\partial_\eta
  O^+_{\xi\alpha} \right) + \frac{1}{3}\left( Q^{(5)}_{\xi,\eta\alpha}
  - \frac{1}{2}\,\partial_\xi O^+_{\eta\alpha} \right)\nonumber\\
& & {} + \bar u i \left[ \dnabla_\xi
  \sigma_{\eta\rho} g(G+i\gamma_5 \wt{G})_{\alpha\rho} - \dnabla_\eta
  \sigma_{\xi\rho} g(G+i\gamma_5 \wt{G})_{\alpha\rho}\right] d,
\end{eqnarray}
where $O^+_{\xi\eta} = \bar u (G+i\gamma_5 \wt{G})_{\xi\eta} d$.
The matrix element of the first three terms on the right-hand side is:
\begin{equation}
\langle 0 | -\frac{2}{3}\, (\dots) - \frac{1}{3}\, (\dots) +
\frac{1}{3}\, (\dots) | \rho^-\rangle = P_\alpha (e_\eta P_\xi - e_\xi
P_\eta) f_\rho^T m_\rho^2 \left\{ 
\langle\!\langle Q^{(5)}\rangle\!\rangle  - \frac{1}{2}\, (\zeta_4^T +
  \wt\zeta_4^T) \right\}.
\end{equation}
For the remaining term on the right-hand side we use that
$$
\sigma_{\eta\rho} i\gamma_5 \wt{G}_{\alpha\rho} = \sigma_{\alpha\rho}
G_{\eta\rho} - \frac{1}{2}\, g_{\alpha\eta} \sigma G,
$$
and observe  that the term in $\sigma G$ has zero matrix element over the
$\rho$ meson.
Thus, using (\ref{eq:star}), we find
\begin{equation}
\langle 0 | \bar u i \dnabla_\xi
  (\sigma_{\eta\rho} gG_{\alpha\rho} + \sigma_{\alpha\rho}
  gG_{\eta\rho}) d - (\xi\leftrightarrow\eta) | \rho^-\rangle =
  P_\alpha (P_\xi e_\eta - P_\eta e_\xi) [2A+2B-4C-3D-3E].
\end{equation}
This gives
\begin{equation}
-\langle\!\langle Q^{(3)}\rangle\!\rangle 
= 2A+2B-4C-3D-3E + \langle\!\langle Q^{(5)} \rangle\!\rangle-
 \frac{1}{2}\,(\zeta_4^T+\wt\zeta_4^T),
\end{equation}
and with $A$, \dots, $E$ from (\ref{eq:LE}), finally:
\begin{equation}
\langle\!\langle Q^{(3)}\rangle\!\rangle 
+ \langle\!\langle Q^{(5)}\rangle\!\rangle 
= \frac{3}{14}\,\zeta_3\omega_3^T + \frac{2}{7}\,
(t_{10} + \wt t_{10}) - \frac{1}{14}\, (\eta_1 + 9\eta_2) -
\frac{1}{14}\, (\wt\eta_1 + 9 \wt\eta_2).\label{eq:B.37}
\end{equation}
The eight relations (\ref{eq:etas}), (\ref{eq:B.18}), (\ref{eq:B.19}),
(\ref{eq:B.23}), (\ref{eq:B.25}), (\ref{eq:xyz}), (\ref{eq:B.32}),
(\ref{eq:B.37}) yield $s_{10}$, $s_{01}$, $\wt s_{10}$, $\wt
s_{01}$, $t_{10}$, $\wt t_{10}$ as given in (\ref{eq:CO}).

\section{Numerical Estimates:  QCD Sum Rules
         }\label{app:C}
\setcounter{equation}{0}

In this appendix we estimate
the independent nonperturbative parameters from QCD sum rules.
The sum rules for chiral-odd
matrix elements can be adapted from the analysis of the photon
distribution amplitudes in Ref.~\cite{BBK}, while those for chiral-even
matrix elements are partly available from Ref.~\cite{BK86}, partly new. The
numerical results are collected in Tables~\ref{tab:para1}, 
\ref{tab:para2} and \ref{tab:para3}.

\subsection{Chiral-Even}

To leading conformal spin accuracy,  we need the single 
parameter $\zeta_4$ (\ref{def:zeta4}). 
This matrix element was discussed at great length in \cite{BK86}.
The best estimate comes from considering the correlation function
\begin{equation}
i\int d^4x\, e^{iqx}\,\langle 0 | T\bar
u(x)g\widetilde{G}_{\mu\alpha}\gamma^\alpha\gamma_5 d(x)\: \bar d(0)
\gamma_\nu u(0) | 0\rangle =
 (q_\mu q_\nu - g_{\mu\nu} q^2) \Pi_{\zeta_4}(q^2),
\end{equation}
which yields the following  sum rule \cite{BK86}:
\begin{eqnarray}
\lefteqn{f_\rho^2 \,\frac{m_\rho^2}{M^2}\,\zeta_4 e^{-m_\rho^2/M^2}
  =}\nonumber\\
& = &
-\frac{\alpha_s}{18\pi^3}\, M^2 \left\{ 1-e^{-s_0/M^2} \left( 1 +
    \frac{s_0}{M^2} \right)\right\} + \frac{\langle (\alpha_s/\pi)
  G^2\rangle}{6M^2} - \frac{32}{27M^4}\, \pi\alpha_s\langle\bar q q\rangle^2.
\end{eqnarray}
In the Borel window $1\,{\rm GeV}^2\leq M^2 \leq 2\,{\rm GeV}^2$ and
with $s_0\approx 1.5\,{\rm GeV}^2$ and the condensates 
$\langle (\alpha_s/\pi)  G^2\rangle = (0.012\pm 0.006)\,{\rm GeV}^4$ and
$\langle\sqrt{\alpha_s}\bar q q\rangle^2 = 0.56\cdot(-0.25\,{\rm
  GeV})^6$, we obtain 
\begin{equation}
\zeta_4(\mu = 1\,{\rm GeV}) = 0.15\pm 0.10.
\end{equation}

The calculation of $\omega_4^A$ involves dimension 6 operators for which 
the QCD sum rule approach becomes rather unreliable. Because of this, 
we choose to make a simple estimate by considering the leading 
contribution to a correlation function that vanishes in perturbation
theory:    
\begin{eqnarray}
CF_X & = & i\int\!\! d^4y \, e^{iqy} \langle 0 | T \bar
d(y)\sigma_{\kappa\lambda} u(y) O^{(1)}_{\xi\beta\xi\nu}(0) |
0\rangle\nonumber\\
& = & \frac{i}{2}\,\langle \bar q
\sigma gG q\rangle\,\frac{q_\beta}{q^2} (q_\kappa
g_{\lambda\nu} - q_\lambda g_{\kappa\nu} ) + O(1/q^4).\label{eq:CFa1}
\end{eqnarray}
The contribution of the $\rho$ meson to this correlation function is
\begin{equation}
CF_X = \frac{i}{m_\rho^2-q^2} \,f_\rho f_\rho^T m_\rho^3
\left[ q_\nu (q_\kappa g_{\lambda\beta} - q_\lambda g_{\kappa\beta} )
  (X_+^{(1)} + X_-^{(1)}) + q_\beta (q_\kappa
g_{\lambda\nu} - q_\lambda g_{\kappa\nu} ) (X_+^{(1)} - X_-^{(1)}) \right],
\end{equation}
from which we obtain in the local duality limit $q^2\to-\infty$:
\begin{eqnarray}
X_+^{(1)}  \simeq -  X_-^{(1)} & \simeq
 & -\frac{1}{4f_\rho f_\rho^T m_\rho^3}\,
\langle \bar q \sigma gG q\rangle
\end{eqnarray}
at a hadronic scale $\mu\approx 1\,$GeV. 
This has to be compared with the estimate for $\zeta_4$ obtained in
the same approximation by considering a similar correlation function 
with the operator (\ref{def:zeta4}):
\begin{eqnarray}
 \zeta_4 & \simeq & -\frac{1}{2f_\rho f_\rho^T m_\rho^3}\, 
\langle \bar q \sigma gG q\rangle.
\end{eqnarray}
Putting in numbers, we get $\zeta_4\approx 0.3$, which is a factor two
larger than what comes from the more accurate 
(and laborious) analysis in \cite{BK86}, see above.
Using the definition (\ref{eq:X}) we get, finally
\begin{eqnarray}
 \omega_4^A(1\,{\rm GeV}) \approx 7/9
\end{eqnarray}
with, probably, a 100\% error.

\subsection{Chiral-Odd}

The calculation of the matrix elements $\zeta_4^T$ and
$\widetilde{\zeta}_4^T$, defined in (\ref{eq:defzetaT}), and the
matrix elements $ \langle\!\langle Q^{(i)}\rangle\!\rangle$ of the
operators $Q^{(i)}_{\alpha,\xi\eta}$, $i = 1,3,5$, 
defined in (\ref{eq:defmeQ}), is analogous to calculation of
the parameters of the photon distribution function in Ref.~\cite{BBK},
and the sum rules obtained in this paper can be adapted to the 
present case.

To leading conformal spin accuracy, we need to estimate two 
parameters, $\zeta_4^T$ and $\widetilde{\zeta}_4^T$.
To this end, we consider the correlation functions
\begin{equation}
CF_\pm = i\int d^4y \, e^{iqy} \langle 0 | T \bar d(y) \gamma_\mu
u(y)\, \bar u(0) g[G(0)\pm i \gamma_5 \widetilde{G}(0)]_{\alpha\beta}
d(0) | 0\rangle\,,
\end{equation}
which vanish in perturbation theory. The leading power corrections
were calculated in \cite{BBK}, yielding
\begin{eqnarray}
CF_+ & = & i(q_\beta g_{\mu\alpha} - q_\alpha g_{\mu\beta} )\,
O(1/q^4),\nonumber\\
CF_- & = & i(q_\beta g_{\mu\alpha} - q_\alpha g_{\mu\beta} ) \left\{
 \frac{\langle \bar q \sigma gG q\rangle}{3q^2} + O(1/q^4)\right\}.
\end{eqnarray}
Saturation with a $\rho$ meson gives, on the other hand,
\begin{equation}
CF_\pm = \frac{1}{m_\rho^2-q^2}\, m_\rho^3f_\rho f_\rho^T (\zeta_4^T \pm
\widetilde{\zeta}_4^T ) i (g_{\mu\alpha} P_\beta - g_{\beta\mu}
P_\alpha),
\end{equation}
so that, taking into account only the leading $1/q^2$ terms, we have
\begin{eqnarray}
\zeta_4^T + \widetilde{\zeta}_4^T & = & 0,\nonumber\\
\zeta_4^T - \widetilde{\zeta}_4^T & = & -\frac{1}{3}\, \frac{\langle
  \bar q \sigma gG q\rangle}{m_\rho^3 f_\rho f_\rho^T}.
\end{eqnarray}
Using the same numerical input as in the last section, we obtain
\begin{equation}
\zeta_4^T(1\,{\rm GeV}) = -\widetilde{\zeta}_4^T(1\,{\rm GeV}) = 0.10\pm 0.05
\end{equation}
with a rather conservative large error.

We use the same method to estimate also the $\langle\!\langle
Q^{(i)}\rangle\!\rangle$, and consider the correlation functions
\begin{equation}
CF^{(i)} = i\int d^4y \, e^{iqy} \langle 0 | T \bar d(y) \gamma_\kappa
u(y)\, Q^{(i)}_{\alpha,\xi\eta}(0)| 0 \rangle.
\end{equation}
As shown in \cite{BBK}, the lowest order power correction to
$CF^{(3,5)}$ vanishes, so that 
\begin{equation}
 \langle\!\langle Q^{(3)}\rangle\!\rangle = 
\langle\!\langle Q^{(5)}\rangle\!\rangle =0
\end{equation}
to that accuracy. For $ \langle\!\langle Q^{(1)}\rangle\!\rangle$, on the
other hand, the mixed condensate gives a nonzero contribution and we
obtain
\begin{equation}
\langle\!\langle Q^{(1)}\rangle\!\rangle(1\,{\rm GeV}) \simeq \frac{5}{9}\, 
\frac{\langle \bar q \sigma gG q\rangle(1\,{\rm GeV})}{m_\rho^3 
f_\rho f_\rho^T} \simeq -0.30
\end{equation}
This value is likely to be overestimated since the mass scale
in the correlation function is much larger than the $\rho$ meson
mass, see the discussion in \cite{BBK}. We thus prefer to give
\begin{equation}
\langle\!\langle Q^{(1)}\rangle\!\rangle(1\,{\rm GeV}) = -0.15\pm 0.15
\end{equation}  
as a conservative estimate.

\end{document}